\documentclass[iop, tighten, numberedappendix, letterpaper]{emulateapj}

\usepackage{amssymb,amsmath,mathrsfs,graphicx}
\usepackage{verbatim}
\usepackage{subfigure}
\usepackage{url}

\def\h2{$\rm H_2$}

\newcommand{\msun}{M$_{\odot}$}

\renewcommand{\star}{\ast}

\newcommand{\Mmax}{M_{\max}}
\newcommand{\Mlim}{M_{\mathrm{lim}}}
\newcommand{\Mcomp}{M_{\mathrm{comp}}}
\newcommand{\Mmin}{M_{\mathrm{min}}}
\newcommand{\Mmaxobs}{M_{\mathrm{max},\mathrm{obs}}}
\newcommand{\Mmaxinput}{M_{\mathrm{max},\mathrm{input}}}

\newcommand{\lnn}{\mathrm{ln~}}
\newcommand{\Nstar}{N_{\star}}
\newcommand{\Npred}{N_{\mathrm{pred}}}

\newcommand{\midline}{\, | \,}
\newcommand{\valpha}{\vec{\alpha}}
\newcommand{\vtheta}{\vec{\theta}}
\newcommand{\vd}{\vec{d}}

\newcommand{\pMFup}{p_{\mathrm{MF}}( M \midline \vec{\theta})}

\newcommand{\pMFobs}{p_{\mathrm{MF_o}}( M \midline \vec{\theta}, \mathrm{obs})}
\newcommand{\pobsM}{p( \mathrm{obs} \midline M)}
\newcommand{\pobsMbar}{p( \mathrm{obs} \midline \overline{M})}
\newcommand{\pobsd}{p( \mathrm{obs} \midline \vec{d_i})}
\newcommand{\pobstheta}{p( \mathrm{obs} \midline \vec{\theta})}
\newcommand{\pMFi}{p_{\mathrm{MF_o}}( M_i \midline \vec{\theta}, \mathrm{obs}_i)}
\newcommand{\pMlikely}{p_\mathrm{M} (\{M_i\} \midline \vec{\theta}, \{\mathrm{obs}_i\})}
\newcommand{\Mbar}{\overline{\mathrm M}}

\let\oldsqrt\sqrt
\def\sqrt{\mathpalette\DHLhksqrt}
\def\DHLhksqrt#1#2{%
\setbox0=\hbox{$#1\oldsqrt{#2\,}$}\dimen0=\ht0
\advance\dimen0-0.2\ht0
\setbox2=\hbox{\vrule height\ht0 depth -\dimen0}%
{\box0\lower0.4pt\box2}}

\begin{document}

\shortauthors{Weisz et al.}
\title{The Panchromatic Hubble Andromeda Treasury IV.  A Probabilistic Approach to Inferring the High Mass Stellar Initial Mass Function and Other Power-law Functions}

\thanks{Based on observations made with the NASA/ESA Hubble Space Telescope, obtained from the Data Archive at the Space Telescope Science Institute, which is operated by the Association of Universities for Research in Astronomy, Inc., under NASA constract NAS 5-26555.}

\author{
Daniel R.\ Weisz\altaffilmark{1},
Morgan Fouesneau\altaffilmark{1},
David W.\ Hogg\altaffilmark{2,3},
Hans-Walter Rix\altaffilmark{3},
Andrew E.\ Dolphin\altaffilmark{4}, 
Julianne J.\ Dalcanton\altaffilmark{1},
Daniel T.\ Foreman-Mackey\altaffilmark{2},
Dustin Lang\altaffilmark{5},
L. Clifton Johnson\altaffilmark{1},
Lori C. Beerman\altaffilmark{1},
Eric F.\ Bell\altaffilmark{6},
Karl D.\ Gordon\altaffilmark{7},
Dimitrios Gouliermis\altaffilmark{3,8},
Jason S.\ Kalirai\altaffilmark{7},
Evan D.\ Skillman\altaffilmark{9},
Benjamin F. Williams\altaffilmark{1}
}

\altaffiltext{1}{Department of Astronomy, University of Washington, Box 351580, Seattle, WA 98195, USA; dweisz@astro.washington.edu}
\altaffiltext{2}{Center for Cosmology and Particle Physics, New York, University, 4 Washington Place, New York, NY 10003, USA}
\altaffiltext{3}{Max Planck Institute for Astronomy, Koenigstuhl 17, 69117 Heidelberg, Germany}
\altaffiltext{4}{Raytheon Company, 1151 East Hermans Road, Tucson, AZ 85756, USA}
\altaffiltext{5}{Department of Astrophysical Sciences, Princeton University, Princeton, NJ 08544, USA}
\altaffiltext{6}{Department of Astronomy, University of Michigan, 500 Church St., Ann Arbor, MI 48109, USA}
\altaffiltext{7}{Space Telescope Science Institute, 3700 San Martin Drive, Baltimore, MD, 21218, USA}
\altaffiltext{8}{Universit\"at Heidelberg, Zentrum f\"ur Astronomie, Institut f\"ur Theoretische Astrophysik,
Albert-Ueberle-Str.~2, 69120 Heidelberg, Germany}
\altaffiltext{9}{Minnesota Institute for Astrophysics, University of Minnesota, 116 Church Street SE, Minneapolis, MN 55455, USA}

\begin{abstract}

 We present a probabilistic approach for inferring the parameters of
 the present day power-law stellar mass function (MF) of a resolved young star cluster. This
 technique (a) fully exploits the information content of a given
 dataset; (b) can account for observational uncertainties in a
 straightforward way; (c) assigns meaningful uncertainties to the
 inferred parameters; (d) avoids the pitfalls associated with
 binning data; and (e) can be applied to virtually any resolved young
 cluster, laying the groundwork for a systematic study of the high
 mass stellar MF (M $\gtrsim$ 1 \msun).  Using simulated clusters and Markov chain Monte
 Carlo sampling of the probability distribution functions, we show that estimates of the MF slope, $\alpha$, are unbiased and that the uncertainty, $\Delta \alpha$,  depends primarily on the number of observed stars and on the range of stellar masses they span, assuming that the uncertainties on individual masses and the completeness are both well-characterized.    
 Using idealized mock data, we compute the theoretical precision, i.e., lower limits, on $\alpha$, and provide an analytic approximation for  $\Delta \alpha$ as a function of the observed number of stars and mass range.  Comparison with literature studies shows that $\sim$ 3/4 of quoted uncertainties are smaller than the theoretical lower limit.  By correcting these uncertainties to the theoretical lower limits, we find the literature studies yield $\langle \alpha \rangle =$ 2.46, with a 1-$\sigma$ dispersion of 0.35 dex. We verify that it is impossible for a power-law MF to obtain meaningful constraints on the  upper mass limit of the IMF, beyond the lower bound of the most massive star actually observed. We show that avoiding substantial biases in the MF slope requires: (1) including the MF as a prior when deriving individual stellar mass estimates; (2) modeling the uncertainties in the individual stellar masses; and (3) fully characterizing and then explicitly modeling the completeness for stars of a given mass.  The precision on MF slope recovery in this paper are lower limits, as we do not explicitly consider all possible sources of uncertainty, including dynamical effects (e.g., mass segregation), unresolved binaries, and non-coeval populations.  We briefly discuss how each of these effects can be incorporated into extensions of the present framework. Finally, we emphasize that the technique and lessons learned are applicable to more general problems involving power-law fitting.

\end{abstract}


\section{Introduction}
\label{sec:intro}

The high mass end of the stellar initial mass function (IMF) underpins much of extragalactic astrophysics. Stars more massive than a few solar masses are largely responsible for most chemical enrichment, dominate the spectral energy distribution blueward of $\sim$ 1$\mu$m for all star-forming galaxies, and are presumed to dominate stellar feedback processes on galactic scales. Consequently, the exact numbers and mass distributions of high mass stars are central to the interpretation of integrated light from distant galaxies, chemical evolution models, the frequency of core-collapse supernovae, the evolution of star formation rates over cosmic time, and the efficiency of star formation on galactic scales \citep[e.g.,][]{lei99, bru03, tin80, sma09, mad96, sch59, ken89, ler08}.  Further, the form of the IMF and its potential sensitivity to local environment carries significant implications for how stars form from dense molecular cores and how the individual stars affect the properties of stellar clusters  \citep[e.g.,][]{elm99, elm04, wei05, zin07, mck07, wei10, por10, bas10}.  

The IMF can be sensibly parameterized in a number of ways \citep[e.g.,][]{sal55, mil79, sca86, cha03}, in particular as a set of power-laws.  Following \citet{kro01}, we adopt the following parameterization of the IMF: 

\begin{equation}
\Phi(M) \equiv \langle \frac{dN}{dM} \rangle = c_i \, M^{-\alpha_{i}} \,\ , \;\   M_{a,i} \le M \le M_{b,i} \, ,
\label{eq:imf}
\end{equation}

\noindent where $c_{i}$ are chosen to ensure continuity, $\Phi(M)$ is normalized such that

\begin{equation}
\int \Phi(M) \, \mathrm{d}M = 1 \, M_{\odot} \, ,
\label{eq:imfnorm}
\end{equation}

\noindent and  $\valpha\equiv \alpha_{i}, i=1,3$ is slope of each power-law component within specified mass interval such that

\begin{align}
\alpha_{1} &=  0.3, \;\ 0.01 \le M/M_{\odot} < 0.08 \nonumber   \\
\alpha_{2} & =  1.3, \;\   0.08 \le M/M_{\odot} < 0.5 \nonumber \\
\alpha_{3} & =  2.3, \;\    0.5 \le M/M_{\odot} \le \Mmax \, , \nonumber 
\end{align}
\label{eq:kroupa}

\noindent with $\Mmax$ being the maximal mass of a star that could have been formed. The value of $\Mmax$ can either be set by the maximal mass at which the stellar lifetime exceeds the (finite) age of the cluster, or by the star- or cluster- formation process itself \citep[e.g.,][]{elm02, elm04, wei05, goo09, mye10, kru10, kru11}.  The high mass portion of the IMF (i.e., $\alpha_3$) sets essentially all observables (e.g., luminosity, color, chemical enrichment, etc.) in extragalactic contexts, making analysis of a single power-law a reasonable approximation in most environments outside the Galaxy.  We will therefore used shorthand notation $\alpha \equiv \alpha_3$ throughout this paper.

\begin{figure}[t]
\begin{center}
\epsscale{1.0}
\plotone{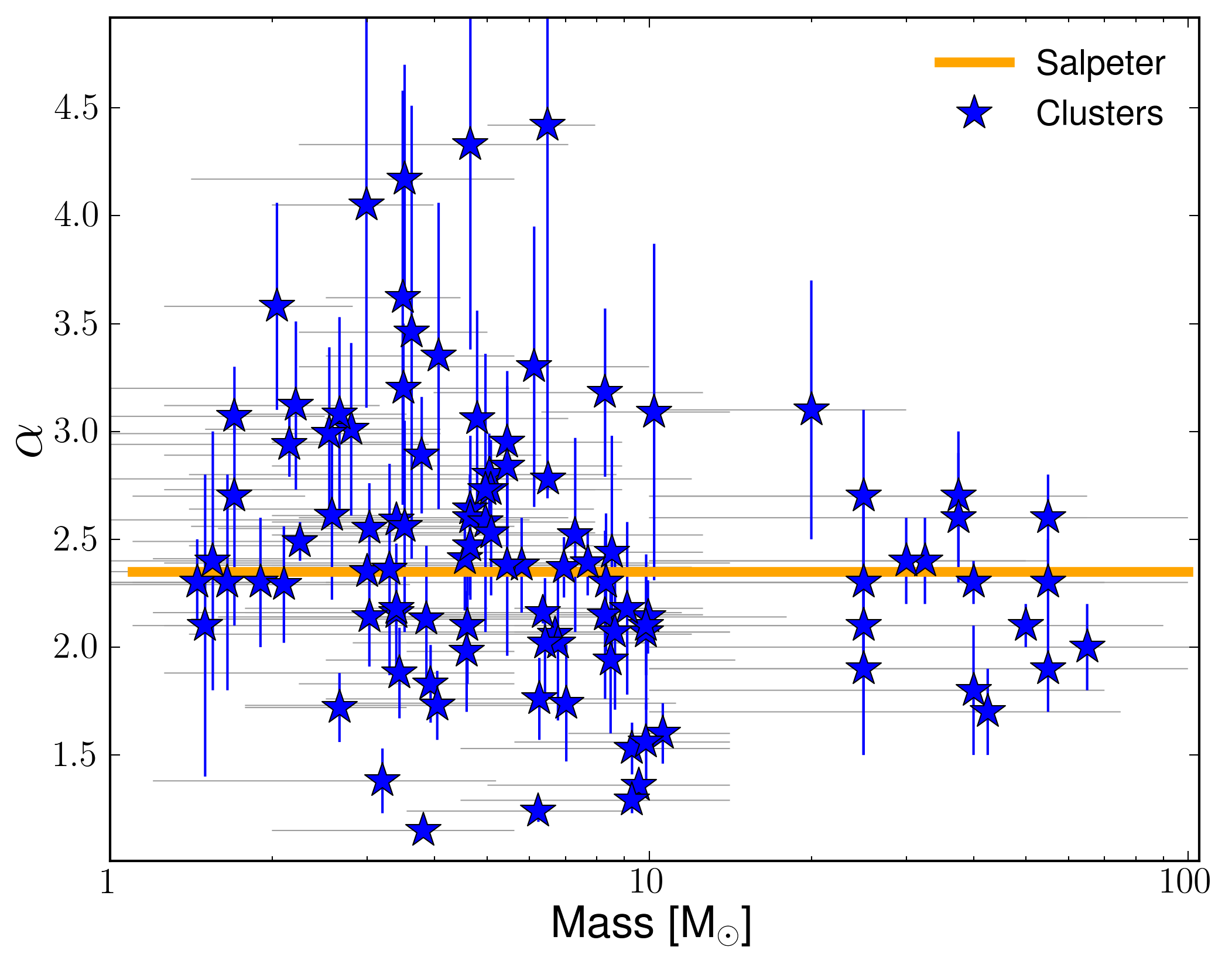}
\caption{Selected literature measurements for the high mass ($M$ $\gtrsim$ 1 \msun) IMF slope from resolved star counts (see Table \ref{tab1}). Points have been placed on the x-axis at the mean mass of the range covered by a given study.  The thin grey lines indicate the span of the full mass range.  Error bars in the y-direction reflect the quoted 1-$\sigma$ uncertainties. The heterogeneity in the MF recovery techniques makes it difficult to discern whether the $\sim$ 0.5 dex of scatter of the data is due to physical or observational challenges and further convolutes the significance of the uncertainties. }
\label{fig:bastian_fig2}
\end{center}
\end{figure}

Despite its widespread importance, the IMF for stars above $\sim$ 1 \msun\ remains insufficiently constrained by observations.  As shown in Figure \ref{fig:bastian_fig2} and listed in Table \ref{tab1}, measurements of the IMF slope from studies of resolved star clusters exhibit a scatter of $\sim$ 0.5 dex.  Additionally, various compilation `$\alpha$-plots' from the literature derive mean IMF slopes with variations of $\sim$ 0.4 dex between different compilations \citep[e.g.,][]{sca98, mar10, bas10}. Taken at face value, the observed scatter in individual IMF measurements and variations in the average values of different compilations introduce significant systematics into interpreting the higher redshift universe \citep[e.g.,][]{coo08, con09, nar12}, and further inhibit differentiation among star formation models that result in a universal IMF \citep[e.g.,][]{mck07, mye10} and those that are sensitive to environmental conditions \citep[e.g.,][]{wei05}, each of which present a drastically different view of how galaxies evolve \citep[e.g.,][]{pfl09, haa10}.

However, at present, it remains unclear whether the precision and accuracy of IMF slope measurements  should be taken at face value, which either implies a formal rejection of a universal high mass IMF slope or whether it should be interpreted as a reflection of stochasticity in the measurements. 

Currently, the majority of high mass IMF studies directly count stars or measure luminosity functions in single or small sets of clusters, with each employing different observational and IMF recovery techniques.    Many studies selectively address critical issues including corrections for binary stars, spatially dependent observational completeness, and uncertainties in individual stellar masses. Consequently, whether the broad range of measurements in Figure \ref{fig:bastian_fig2} are due to intrinsic physical effects or are the result of systematics remains an open and important question (see discussions in \citealt{sca86, elm99, kro02, sca06, elm09, bas10}).

In practice, we can only measure the present day mass function (MF) of evolved clusters, in which the observable stellar masses are limited at the high mass end by the turnoff mass corresponding to the age of the cluster, although the object of ultimate importance is the entire IMF.  For a simple stellar population, i.e., a star cluster, the measured MF slope is identical to the IMF slope \citep[e.g.,][]{elm06} modulo certain dynamical (e.g., mass segregation) and observational (e.g., completeness) effects. Consequently, our understanding of the IMF critically depends on our ability to accurately and precisely constrain the MF.

Systematically counting the relative numbers of individual stars (with $M \gtrsim$ 1 \msun) in a large collection of young clusters has long been advocated as the optimal approach to constraining the high mass MF \citep[e.g.,][]{sca86, kro02, sca06, bas10, kro11}. As improved datasets of resolved stars have emerged \citep[e.g.,][]{zar97, mas06, hol06, dal09, dal12, bia12},  it is crucial that the analysis tools for constraining the high mass MF are in place to fully exploit the information content of the data and to properly account for all known uncertainties.   

Broadly speaking, the practical procedure of obtaining a MF constraint for a resolved young cluster consists of three steps. First, one obtains individual mass estimates for the stars, either based on photometry in conjunction with isochrones, or from spectroscopy. Second, one determines the completeness of the sample for which mass estimates were obtained. Finally, one fits a MF model (e.g., Equation \ref{eq:imf}) to the data to obtain confidence limits on the slope parameter, $\alpha$. In its usual implementation, this approach uses $\chi^2$ minimization of a power-law model fit to a binned or cumulative representation of the mass function, $N(M)$. The fit is taken over the masses determined to be sufficiently complete and the error bars on each mass bin reflect Poisson statistics of the number of stars in the bin. 

However, there are severe limitations in this traditional approach.  Correlations between the number of stars per bin and the adopted bin size can introduce significant biases into the resulting MF slope and associated uncertainties \citep[e.g.,][]{mai05, car08}.  Further, such an approach does not provide an objective framework for folding in observational effects such as completeness and mass uncertainties, which are critical to accurate MF constraints.  More robust approaches such as Monte-Carlo simulations, color-magnitude diagram fitting, non-parametric kernel estimators, and maximum likelihood techniques \citep[e.g.,][]{els89, gen11, dol02, vio94, tar82} alleviate some of these issues.  However, their reliability has not been shown in large, diverse sets of clusters, the interpretation of the error bars is often ambiguous, and there remain statistical short-comings in even the most advanced conventional approaches \citep[e.g., noise amplification in maximum likelihood estimations; ][]{gou97}. In short, traditional approaches to MF reconstruction are inadequate for a systematic investigation of the high mass stellar IMF.

The main goal of this paper is to lay out a forward modeling technique that remedies at least some of the shortcomings of the traditional approach. It should result in constraints on the high mass MF that come close to exploiting the full information content of the data and that explicitly account for known observational uncertainties. In particular, by laying out a forward modeling formalism, we aim to overcome the following limitations of many previous analyses by: (1) accounting for the mass estimate errors of individual stars that may arise from the conversion of observed fluxes into stellar masses; (2) avoiding binning the measurement in mass, as the most massive bins are unavoidably sparsely populated; (3) dealing explicitly with completeness functions that may not be a simple step function of stellar mass, and instead be either gradual or varying as a function of position; (4) implementing a rigorous way to derive the joint posterior distribution function of the MF parameters (e.g., $\alpha$ and $\Mmax$) in light of evidence for the data and its uncertainties and possibly other prior information. 

In this paper, we present the probabilistic framework for inferring the parameters of the high mass MF of a young stellar cluster.  We carefully walk though the necessary mathematics and interlace illustrative examples using simulated clusters.   For clarity in the derivations and examples, we have made several simplifying assumptions.  In particular, dynamical effects such as mass segregation, ejection, and relaxation are not included in the present paper, although their role in interpreting the observed MF as the IMF is indisputably important.   

The methodology presented in this paper is motivated in part by the Panchromatic Hubble Andromeda Treasury program \citep[PHAT;][]{dal12}.  This Hubble Space Telescope multi-cycle treasury program is mapping $\sim$ 1/4 of M31's star forming disk, providing near-UV through near-IR imaging of $\sim$ 10$^8$ resolved stars.  Based on analysis of the Year 1 data, we anticipate the survey will resolve $\gtrsim$ 1000 young clusters \citep[i.e., ages $\lesssim$ 100 Myr; or a main sequence upper mass limit of $\gtrsim$ 5 \msun;][]{joh12} with sensitivity down to stars with M $\sim$ 1-3 \msun, which will enable a large scale study of the high mass stellar MF.  Although the terminology `high mass' is often reserved for stars with $M$ $\gtrsim$ 10-15 \msun, common IMF parameterizations (cf. Equation \ref{eq:kroupa}) suggest that all stars above $\sim$ 1 \msun\ are drawn from the same underlying IMF, indicating that conventionally defined intermediate and massive stars can be used to learn about their common IMF.  Throughout the paper, we use these numbers as guides for our simulations, but emphasize that the methodology developed in this paper can readily be generalized to measure the MF of any resolved cluster and for any parameterized MF model.  

The remainder of this paper is organized as follows. In \S \ref{sec:probimf}, we lay out the general probabilistic framework for inferring the MF of a young stellar cluster.  We then apply this technique to highly idealized mock data (e.g., no mass uncertainties, perfectly known completeness) and present illustrative results in \S \ref{sec:idealdata}.  In \S \ref{perfectdata} and \S \ref{sec:complit}, we derive the theoretical lower limits for recovered MF slope precision and compare the results to select literature studies.  We then consider the more complex case of masses with finite mass uncertainties in \S \ref{sec:massunc}, and explore the case where the stellar mass distributions are log-normal functions in \S \ref{sec:loggaussmass}.  We then consider the case of linear completeness functions in \S \ref{sec:completeness}.  Finally, we discuss several caveats to the MF models adopted in this paper (e.g., the influence of cluster dynamics, non-coeval populations, unresolved binaries) in \S \ref{sec:caveats}.

\section{Probabilistic Framework for Modeling the \\  Stellar Mass Function of a Cluster}
\label{sec:probimf}

\begin{figure}[t]
\begin{center}
\epsscale{1.0}
\plotone{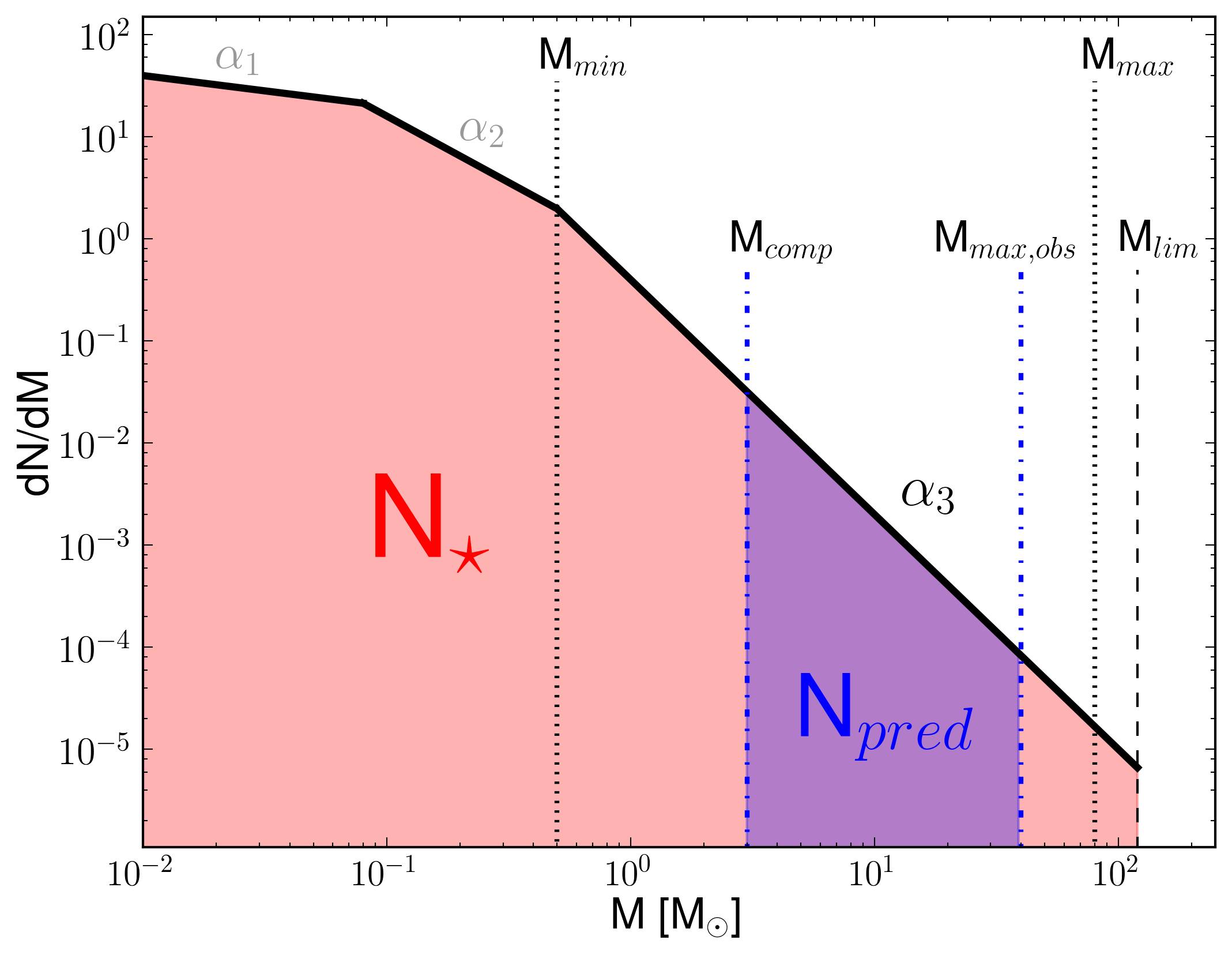}
\caption{A schematic representation of the Kropua IMF over the entire stellar mass spectrum for a hypothetical star cluster (Equation \ref{eq:imf}).  Here, $\Nstar$ (red shading) represents the total number of stars in the cluster and $\Npred$ is the number of stars expected in the observed mass range bounded by $\Mcomp$ and $\Mmaxobs$ (blue shaded region), i.e., the MF.   The upper end of a hypothetical power-law IMF is defined to be between $\Mmin$ and $\Mmax$, which is the maximal mass of a star that could have formed in the cluster.  $\Mlim$ is the upper bound of possible observed stellar masses, as set by stellar physics. The aim of this paper is to outline an approach for inferring the values of $\alpha$, $\Mmax$, and $\Npred$ of a young cluster, given a set of $N$ observed stellar masses.}
\label{fig:kroupa}
\end{center}
\end{figure}

We begin by illustrating the scope of the problem at hand. Consider the IMF of a hypothetical cluster as shown in Figure \ref{fig:kroupa}.  We presume that the high mass IMF regime of this cluster consists of some number of stars, whose mass distribution follows a power-law with a slope of $\alpha$, and whose masses are bounded by $\Mmin$ and $\Mmax$, which is the mass of the most massive star that could have formed in the cluster, and may not be the same for different clusters.  This `zero-age' state of the cluster is the desired dataset to make direct inferences about its stellar IMF.

However, there are several considerations that only allow us to observe the present day MF of this cluster.   First, due to stellar evolution, the most massive star(s) have likely disappeared, and only stars with masses up to $\Mmaxobs$ can be observed at the present day.  Second, observational completeness (due to stellar crowding, for example)  only permits observations of stars more massive than some lower mass limit, $\Mcomp$. Third, the translation of observed flux into stellar mass provides only an estimate of a star's true mass.  That is, the `observed' mass of a star could be an over- or underestimate of a star's true mass. Further, the observed mass of a star is bounded by $\Mcomp$ and $\Mlim$, the maximum mass allowed by stellar evolution models, which may or may not reflect the upper stellar mass limit in nature. Each of these effects must be carefully considered in order to make an accurate measurement of the present day MF of a cluster, a necessary step to make any meaningful statements about a cluster's stellar IMF.

For clarity in exposition, we will begin with the simplest case where the data, $\vd$, consists of a set of $N$ stars with perfectly known stellar masses, $\{M_i\}$, and the completeness function is also perfectly known.  Given the data, $\vd$, we aim to place constraints on the slope of the MF, $\alpha$, the maximum mass of a star that could have formed, $\Mmax$, and the expected number of observed stars, $\Npred$, which normalizes the amplitude of the MF for this cluster (see Figure \ref{fig:kroupa}).  
Throughout this paper, we assume that we have the observed completeness function (either as a function of mass or luminosity) in hand.   In practice, the completeness is a function of both the stellar fluxes (i.e., through the flux limit of the observations) and the local image crowding, and is determined through extensive artificial star tests. Given that most cluster observations, certainly those beyond the Milky Way, are crowding limited, it will always be more challenging to detect and accurately catalog a particular star if it is located near the center of a cluster rather  than in the outskirts.   For the purposes of this paper, we make the simplifying assumption that the completeness function is independent of position.

\subsection{Deriving the General MF Posterior Probability Distribution Function}
\label{sec:genppdf}

We now lay out the general probabilistic framework for constraining the MF of a young cluster. We start by describing the MF of the cluster (cf. Equation \ref{eq:imf}) as:
\begin{equation}
\Phi(M) = N_* \, \pMFup \; ,
\label{eq:MFprob}
\end{equation}

\noindent where $\Nstar$ is the total number of stars in the cluster, $\pMFup$ is the probability distribution function (PDF) for the mass of any star in the cluster, and $M$ is the true mass of the star; the MF is described by the parameters $\vtheta = \{\alpha, \Mmax\}$, where $\alpha$ and $\Mmax$ refer to the high mass portion of the MF, i.e., a single sloped power law such as described by $\alpha_3$ in Equation \ref{eq:imf}.

The PDF for the mass of an observed star, i.e., those that are massive (luminous) enough to end up in our dataset, is given by 
\begin{equation}
\pMFobs \equiv  \frac{\pMFup \, \pobsM}{\pobstheta}\; ,
\label{eqnarray:MFobs}
\end{equation}

\noindent where $\pobsM$ is the probability of observing a star given its true mass $M$, i.e., the completeness function, and $\pobstheta$, the normalization necessary to make $\pMFobs$ a PDF.  This normalization can be written as

\begin{equation}
\pobstheta = \int \pMFup \, \pobsM \, {\mathrm d}M \; .
\label{eqnarray:normMFobs}
\end{equation}

In this model, the expected number of stars in the data list, $\Npred$, sets the amplitude normalization of the MF, and can be expressed as

\begin{equation}
\Npred= N_* \int_0^\infty  \pMFobs \, {\mathrm d}M \; .
\label{eq:nstar}
\end{equation}

However, given that we do not know $\Nstar$ {\it a priori} and that much of the MF is, as in many practical applications, well below our detection limit, we cannot explicitly solve for $\Npred$, and therefore we will treat $\Npred$ as a model parameter that is independent of $\vtheta$.

So far, this model makes predictions for the probability distribution for the mass of any one observed star, $\pMFobs$, for a given set of parameters, $\vtheta$, and for the expected number of observed stars, $\Npred$. Assuming that each of the $N$ observed stars in the cluster is drawn identically and independently from the underlying mass distribution, we can write the probability of measuring the set of $N$ masses, $\{\mathrm M_{i}\}$, as:

\begin{equation}
\pMlikely  = \prod_{i=1}^{N} \pMFi \; .
\label{eq:prodmasses}
\end{equation}

At the same time, we must also include the probability of actually observing $N$ stars in the cluster, as $N$ itself is a datum.  This is the Poisson probability of observing $N$ stars given an expectation of $\Npred$ stars and can be written as

\begin{equation}
p_{\mathrm{Poisson}}(N \midline \Npred) = \frac{\Npred^{N} \, e^{-\Npred}}{N!} \, .
\label{eq:poisson}
\end{equation}

We now have the two expressions that give the probability of the data ($\{M_i\}$, $N$) given model parameters ($\vtheta$, $\Npred$) and the fact that data were observed, i.e., `obs'.  However, we wish to infer the inverse, namely the values of $\vtheta$ and $\Npred$ given a set of $N$ stellar mass measurements.  Using Bayes's theorem, we can then write down the probabilities of the model parameters given the data, i.e., the posterior PDFs (pPDFs) for $\vtheta$ and $\Npred$ as:

\begin{align}
p_{\mathrm{post}}(\vtheta \midline \{M_i\}, \mathrm{obs}) &= \pMlikely \, p_{\mathrm{prior}}(\vtheta) \nonumber \\
&= \prod_{i=1}^{N} \pMFi \, p_{\mathrm{prior}}(\vtheta)  \; ,
\label{eq:posteriortheta}
\end{align}

\noindent and

\begin{align}
p_{\mathrm{post}}(\Npred \midline N) &= p_{\mathrm{poisson}}(N \midline \Npred) \, p_{\mathrm{prior}}(\Npred) \nonumber \\
&= \frac{\Npred^{N} \, e^{-\Npred}}{N!} p_{prior}(\Npred) \; ,
\label{eq:posteriorN}
\end{align}

\noindent where $p_{\mathrm{prior}}(\vtheta)$ and $p_{\mathrm{prior}}(\Npred)$ reflect any prior or extraneous information or constraints on $\vtheta$ and $\Npred$, respectively. In Equations \ref{eq:posteriortheta} and  \ref{eq:posteriorN} we have presumed that the `evidence' terms, $p_{\mathrm{prior}}(\{M_i\})$ and $p_{\mathrm{prior}}(N)$, are both constant, and have omitted them.  In practice, the independence of these two pPDFs allows us to compute them separately, and then simply multiply them together to get the general pPDF for the MF:

\begin{align}
&p_{\mathrm{post}}(\vtheta, \Npred \midline \{M_i \}, N, \mathrm{obs}) = \nonumber \\ &p_{\mathrm{post}}(\vtheta \midline \{M_i\}, \mathrm{obs}) \,  p_{post}(\Npred \midline N)  .
\label{eq:posterior}
\end{align}

The above derivation is limited to a single power-law slope, where the stars are well-above any break or turnover in the MF. Of course, in cases where data on stars well below 1 \msun\ are available, the same formalism carries through, just replacing $\alpha$ with $\valpha$.

\section{Deriving Mass Function Constraints from Idealized Data}
\label{sec:idealdata}

We now show how to explicitly constrain MF parameters given: (1) the pPDF from Equation \ref{eq:posterior}; (2)  evidence for the data; (3) and any prior information on $\vtheta$ and $\Npred$. In practice, the most informative prior on $\vtheta$ frequently comes from $\Mmax$, where astrophysical constraints may exist, especially if there is independent information on the age of the population.  Other general, although less physically informative, priors are requirements that the number of stars is positive and that $\alpha$ does not take on unreasonably extreme values that have been ruled out by previous observations, e.g., $\alpha <$ $-$5.

For the purposes of this initial idealized exercise, we adopt a simple `boxcar' form for the probability of observing a star, given its true mass (i.e., the completeness function), 

\begin{equation}
\pobsM = 
\begin{cases}
1, & 0.5 \, M_{\odot} < \Mcomp \le M \le \Mlim \\
0, & \text{otherwise \; ,}
\end{cases}
\label{eq:simplecompleteness}
\end{equation}

\noindent where $\Mlim$ is the maximum observable mass as set by stellar evolution, $\Mcomp$ is the minimum mass as determined by the observational completeness limit.  For simplicity, we will assume that the MF is a single power-law over the interval $\Mcomp \le M \le \Mlim$ and that $\Mcomp$ is the same value for all locations in the cluster; we discuss the case of spatially varying completeness in \S \ref{sec:caveats}.  In practice, $\pobsM$ is the completeness function of the data.

So far, we have assumed that the uncertainties on masses are negligible.  If follows that a star's observed mass PDF is a delta function, and therefore it is directly interchangeable with a delta function PDF for the true mass of a star.  Consequently, we can use $\pMFobs$ as the generative model for the observations such that

\begin{align}
\pMFobs = \pobsM \, c_{\mathrm{MF_o}}(\vtheta) \, M^{-\alpha}  =  \nonumber \\
\begin{cases}
&c_{\mathrm{MF_o}}(\vtheta) \, M^{-\alpha},  \Mcomp \le M \le \Mmax \\
&0,  \text{otherwise \, ,} 
\end{cases}
\label{eq:probimf}
\end{align}

where the normalization $c_{MF_o}(\vtheta)$ is given by
\begin{equation}
c_{\mathrm{MF_o}}^{-1}(\vtheta) =  \int_{0}^{\infty} \pobsM \, M^{-\alpha}\ dM \, = \int_{\Mcomp}^{\Mmax}M^{-\alpha}\ dM \, .
\label{eq:simplenorm}
\end{equation}

\noindent In this form, the range of the MF model is set by the completeness limit on the low mass end, and the mass of the most massive star to have formed on the upper end.

Using the explicit expressions for each term in Equation  \ref{eq:posterior}, we can now re-write the pPDF for this idealized case as:

\footnotesize
\begin{align}
&p_{\mathrm{post}}(\vtheta, \Npred \midline \{M_{i}\}, N, \mathrm{obs})  = \nonumber \\  &p_{\mathrm{poisson}}(N \midline \Npred) \, \sum_{i=1}^{N} \, \bigl ( c_{\mathrm{MF_o}}(\vtheta) \, M_i^{-\alpha}\bigr ) \,  p_{prior}(\vtheta) \,  p_{\mathrm{prior}}(\Npred) \, ,
\label{eq:longposterior}
\end{align}
\normalsize

In practice we calculate the natural logarithm of the pPDF for computational ease. Also, note that in this idealized limit,  $p_{post}(\vtheta, \Npred | \{M_{i}\}, N, \mathrm{obs})=0$ in the event that $M_i<\Mcomp$ or $\Mmax> \Mlim$.

\subsection{Sampling the the pPDF with a Markov Chain}
\label{mcmc}

To place constraints on the model parameters of interest, we sample the pPDF using a Markov chain Monte Carlo (MCMC) algorithm.  MCMC techniques provide an efficient discrete sampling and clear interpretation of multi-dimensional spaces, such that the density of samples is highest around the most probable parameters and lower in regions of less probable values \citep[e.g.,][]{gel96}.  As a result, MCMC techniques are able to produce well-defined uncertainties for both the one dimensional (marginalized) distributions (e.g., $p(\alpha \midline \{M_i\}, N, \mathrm{obs})$, $p(\Mmax \midline \{M_i\}, N, \mathrm{obs})$)  and reveal degeneracies between two dimensional (joint) distributions (e.g., $p(\alpha, \Npred \midline \{M_i\}, N, \mathrm{obs})$).   An MCMC approach also permits efficient sampling in cases where more parameters are needed to adequately model the MF, e.g., such as when using a multi-component power-law to characterize data containing lower mass stars.

In this analysis, we use the MCMC sampler \texttt{emcee}\footnote{\url{http://danfm.ca/emcee}} as described in \citet{for12}.  $\texttt{emcee}$ is a pure-Python implementation of the affine invariant MCMC ensemble sampler developed by \citet{goo10}.   This sampler explores parameter space through a set of `walkers'.   For each MCMC increment, each walker takes a step in parameter space by choosing another walker and moving along a line in parameter space that connects itself to the other walker.  The size of the step is chosen stochastically, allowing for both interpolation and extrapolation, and the choice in step for each walker is based on the co-variance of the set of walkers. After each step, the pPDF is evaluated for the new set of parameters.  Steps that increase the probability are always accepted, whereas steps that result in a lower probability are {\it sometimes} accepted.  For a sufficiently large number of steps, the ensemble of walkers samples parameter space with a frequency proportional to the pPDF.  In a rough sense, this approach is akin to having a set of parallel Metropolis-Hastings MCMC chains \citep[][]{met53, has70}, but is much more efficient than Metropolis-Hastings sampling, as measured by the autocorrelation time (i.e., the time spent per function call; see discussions in \citealt{goo10}, \citealt{for12}, and the Appendix of this paper). 

$\texttt{emcee}$ requires minimal initial input from the user.  One must select a number of walkers and their initial conditions, and designate the number of burn-in steps and length of each chain.  Following the recommendations in \citet{for12} and experimentation with the simulated datasets, we chose 16 walkers and selected random (valid) values for each walker's starting point.  After testing several combinations, we found that generally 300 burn-in steps and 100 chain steps per walker resulted in stable solutions.  To aide in computational efficiency, we placed sensible restrictions on the allowed ranges for each of the three parameters (i.e., $p_{\mathrm{prior}}(\vtheta)$ and $p_{\mathrm{prior}}(\Npred))$: $-$6 $\le$ $\alpha$ $\le$ 6,  $\Mmaxobs \le \Mmax \le 120$ \msun, and 0 $< \Npred < 10^2\times N$.  In the case of finite mass uncertainties, we revised the restriction on $\Mmax$ for reasons discussed in \S \ref{sec:loggaussmass}.  

We refer the reader to the Appendix for a more in depth explanation of MCMC sampling theory.

\subsection{Illustrative Examples with Idealized Data}
\label{perfectdata}

We now illustrate this approach by applying it to idealized mock data (i.e., no mass uncertainties and a perfectly known boxcar completeness function). A simple way to create the simulated data for a power-law MF, assuming negligible mass uncertainties, is to define a variable $x$ via $M=\exp{\bigl (\ln [x\,(1-\alpha)]/(1-\alpha)\bigr )}$ and then draw a uniform random variate in $x$ between $x_{\mathrm{min/max}, \mathrm{input}}= (1-\alpha)^{-1} \, M^{1-\alpha}_{\mathrm{max/min}, \mathrm{input}}$.  For all illustrative examples in this paper we have adopted $\alpha_{\mathrm{input}} =$ 2.35 \citep[i.e.,][]{sal55}.   We discuss the results of simulations run with other values of $\alpha$ at the end of this section.  

Each dataset consists of $N$ masses that have been discretely sampled over the interval [$M_{\mathrm{min}, \mathrm{input}}$, $\Mmaxinput$].  The value of $M_{\mathrm{min}, \mathrm{input}}$ is equivalent to the lower mass limit, $\Mcomp$, as specified by the completeness function.  For these exercises, we have fixed $M_{\mathrm{min}, \mathrm{input}} =$ 3 \msun, a conservative value for resolved clusters in M31 \citep[e.g.,][]{dal12}.    

The mass of the most massive star that could have formed in a cluster is of great interest astrophysically.   Essentially it provides a clue to the fundamental nature of how stars form, e.g., is there a universal value for the most massive that can form or does it depend on local environmental conditions?  There are several studies in the literature that attempt to quantity the relationship between environment and the mass of the most massive star \citep[e.g.,][]{wei10, cer11, eld12}.  

\begin{figure}[b]
\begin{center}
\plotone{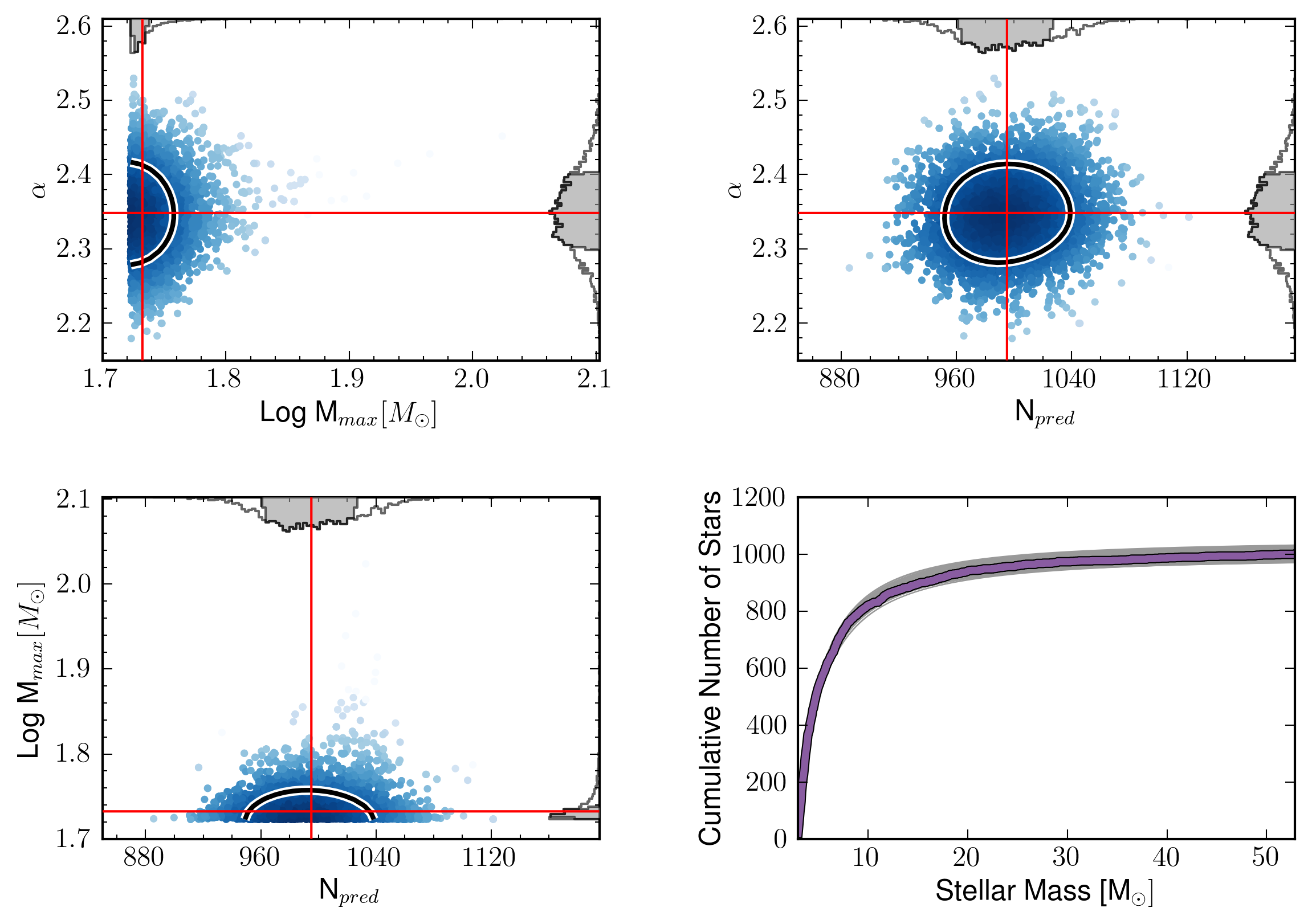}
\caption{The recovery of MF parameters $\alpha$, $\Mmax$, $\Npred$ in the case of idealized mock data, i.e., negligible mass uncertainties and a boxcar completeness function, from a single run of $\mathtt{emcee}$.  The joint distributions are shown as a scatter plot in the center of the first three panels, for the indicated parameters.  Dark colored points are close to the most probable value, while light colors are farther away.  The solid black line encloses 68\% of the points centered around the most probable combination of parameters.  The corresponding marginalized distributions have been projected onto the axis of each plot.  The solid red lines indicate the median of each marginalized distribution, and the shading encloses 68\% of the distribution centered around the median.  The roundness of each joint distribution indicates little degeneracy between the MF parameters.  For this example dataset, we have recovered $\alpha =$ 2.35$_{-0.05}^{+0.05}$, $\Mmax =$ 54.0$_{-0.8}^{+2.2}$, and $\Npred =$ 995$_{-31}^{+33}$. In the lower right hand panel, the observed cumulative mass function is shown as the thick purple line, and 100 MF models randomly drawn from the pPDF are over-plotted as thin grey lines.}
\label{fig:nstar1000}
\end{center}
\end{figure}

In our simulations, the value of the most massive star that could have formed in a cluster is $\Mmaxinput$.  The mass of the most massive star ``observed'', i.e., on the mass list, is $\Mmaxobs$, which is less than $\Mmaxinput$ due to the effects of stellar evolution or dynamical ejection.  Under the assumption that the mass of each star is independently drawn from the IMF,  it is fairly intuitive that $\Mmaxobs$ cannot inform us about $\Mmaxinput$ beyond providing us a simple lower limit on $\Mmaxinput$, i.e., the most massive observed stars does not contain information about more massive stars that are no longer in the cluster.   

For a given dataset, we aim to recover the input MF slope, $\alpha_{\mathrm{input}}$, the maximal mass of a star that could have formed $\Mmaxinput$, and require that our model produce a normalization that is consistent,  i.e., within Poisson uncertainty, with the number of stars observed, $N$.  We denote the corresponding model parameters  as $\alpha$, $\Mmax$, and $\Npred$.  

We applied this technique to mock data for different permutations of $\alpha_{\mathrm{input}}$, $\Mmaxinput$, and $N$.  Results from the first example are shown in Figure \ref{fig:nstar1000}.  Here we show the recovered joint and marginalized distributions for a mock cluster which was generated assuming $N =$ 1000 stars and $\Mmaxinput =$ 60 \msun.  We recover $\alpha =$ 2.35$_{-0.05}^{+0.05}$, $\Mmax =$ 54.0$_{-0.8}^{+2.6}$, and $\Npred =$ 995$_{-31}^{+33}$, where the indicated values represent the median and 0.5$\times$(84th - 16th percentile) of the corresponding marginalized distribution, i.e., 1-$\sigma$ for a Gaussian distribution.  

In general, the recovered values reflect the input values with excellent accuracy and precision.  However, as expected, the recovered value of $\Mmax$ is consistent with the mass of most massive observed star, $\Mmaxobs$, and not that of the most massive star that could have formed $\Mmaxinput$.  Thus, this constraint provides a lower bound on the upper mass limit of the IMF.

The joint distributions in Figure \ref{fig:nstar1000} show little co-variance between parameters, which indicates that there is little degeneracy in this problem, i.e., the estimation of one parameter will not affect the accuracy and precision to which another parameter can be constrained.  Consequently, and for clarity, we will focus on the marginalized distributions for the next set of examples. 

\begin{figure}[t]
\begin{center}
\plotone{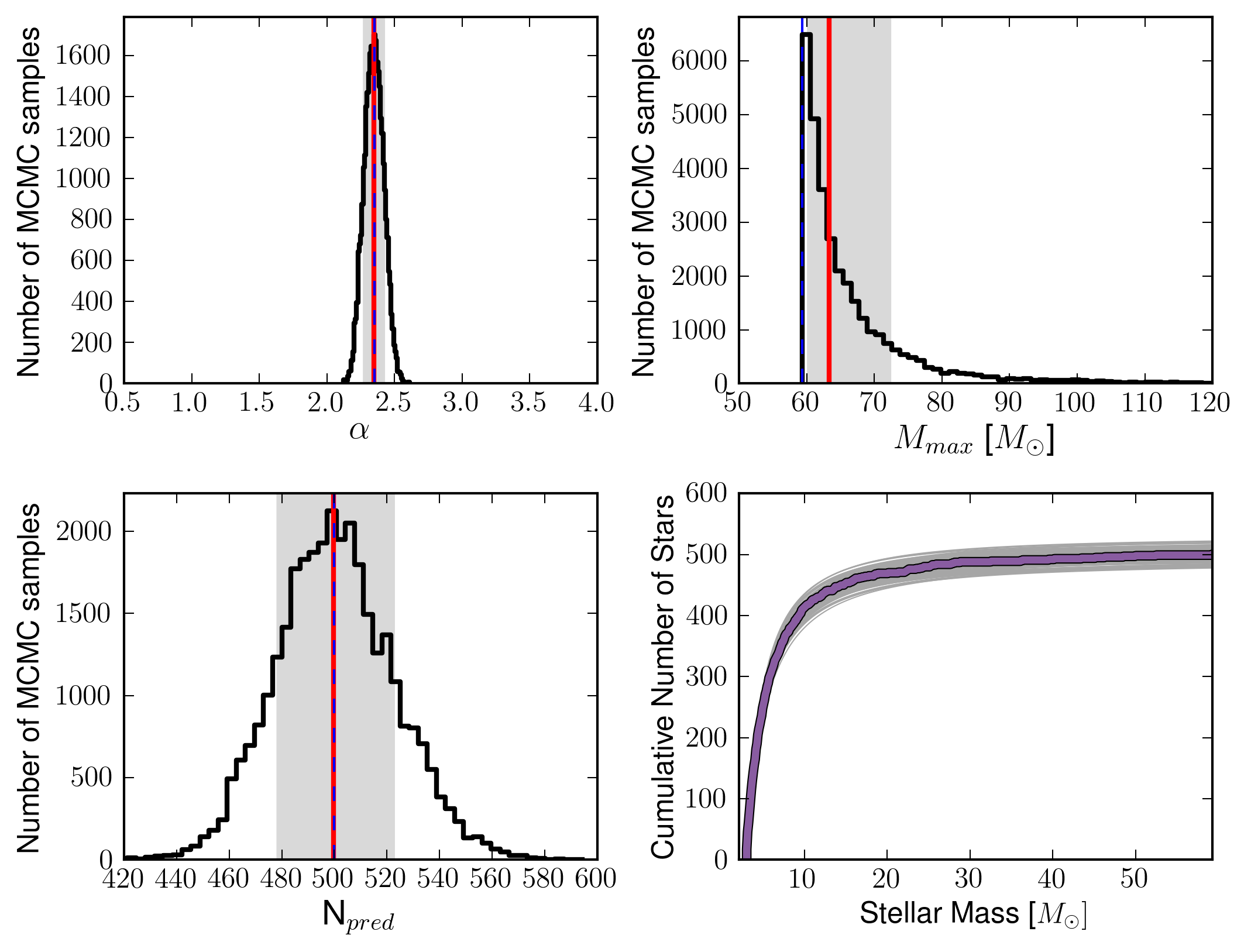}
\caption{The recovery of the MF parameters $\alpha$, $\Mmax$, $\Npred$ in the case of idealized mock data, i.e., negligible mass uncertainties and a boxcar completeness function, from a single run of $\mathtt{emcee}$.  This simulation has $N =$ 500 stars, $\alpha_{\mathrm{input}} =$ 2.35, and $\Mmaxinput =$ 60 \msun.  From top left to lower right: the marginalized distributions for $\alpha$, $\Mmax$, and $\Npred$.  The solid red line represents the median of the distribution and the grey shaded region highlights 68\% of the distribution centered around the median, i.e., the range enclosed by the 16th and 84th percentiles.  In each panel, the blue dashed line represents the `truth', i.e., $\alpha_{\mathrm{input}}$, $\Mmaxobs$, and $N$, respectively.   For this example dataset, we have recovered $\alpha =$ 2.35$_{-0.07}^{+0.07}$, $\Mmax =$ 63.3$_{-3.1}^{+9.0}$, and $\Npred =$ 500$_{-21}^{+22}$, indicating excellent consistency with the input values.  In the case of $\Mmax$, we are only able to place constraints on $\Mmaxobs$ and not on the mass of the most massive star formed.  In the lower right hand panel, the observed cumulative mass function is shown as the thick purple line, and 100 MF models randomly drawn from the pPDF are over-plotted as thin grey lines.}
\label{fig:nstar500}
\end{center}
\end{figure}

\begin{figure}[th!]
\begin{center}
\plotone{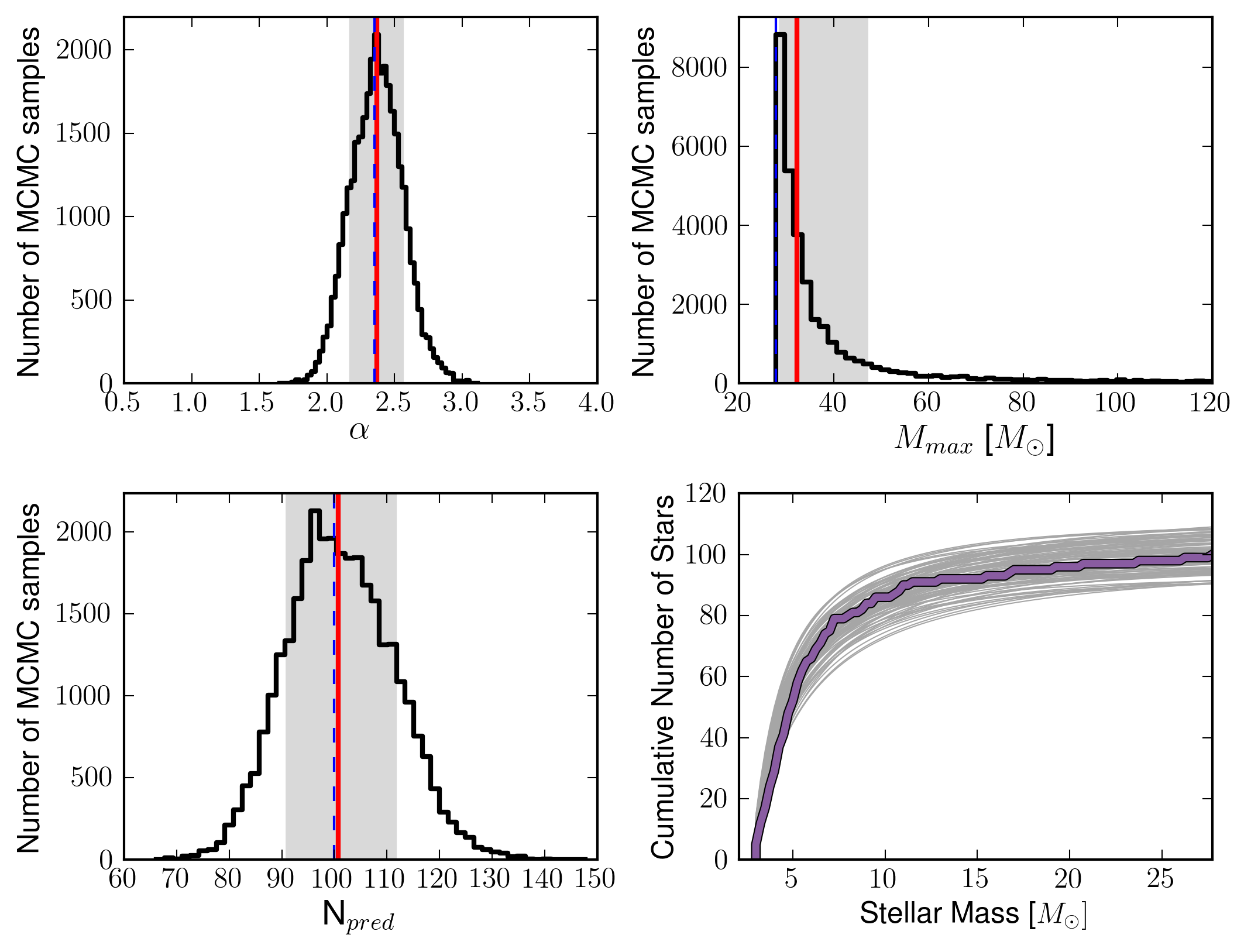}
\caption{The same as Figure \ref{fig:nstar500} only with $N =$ 100 stars and $\Mmaxinput =$ 30 \msun. Here, we recover $\alpha =$ 2.37$_{-0.20}^{+0.19}$, $\Mmax =$ 32.2$_{-3.5}^{+14.8}$ $\Npred =$ 101$_{-10}^{+11}$.  The recovered fractional precision on $\alpha$ is roughly a factor of 2.5 larger than for the example considered in Figure \ref{fig:nstar500}.}
\label{fig:nstar100}
\end{center}
\end{figure}

\begin{figure}[b!]
\begin{center}
\plotone{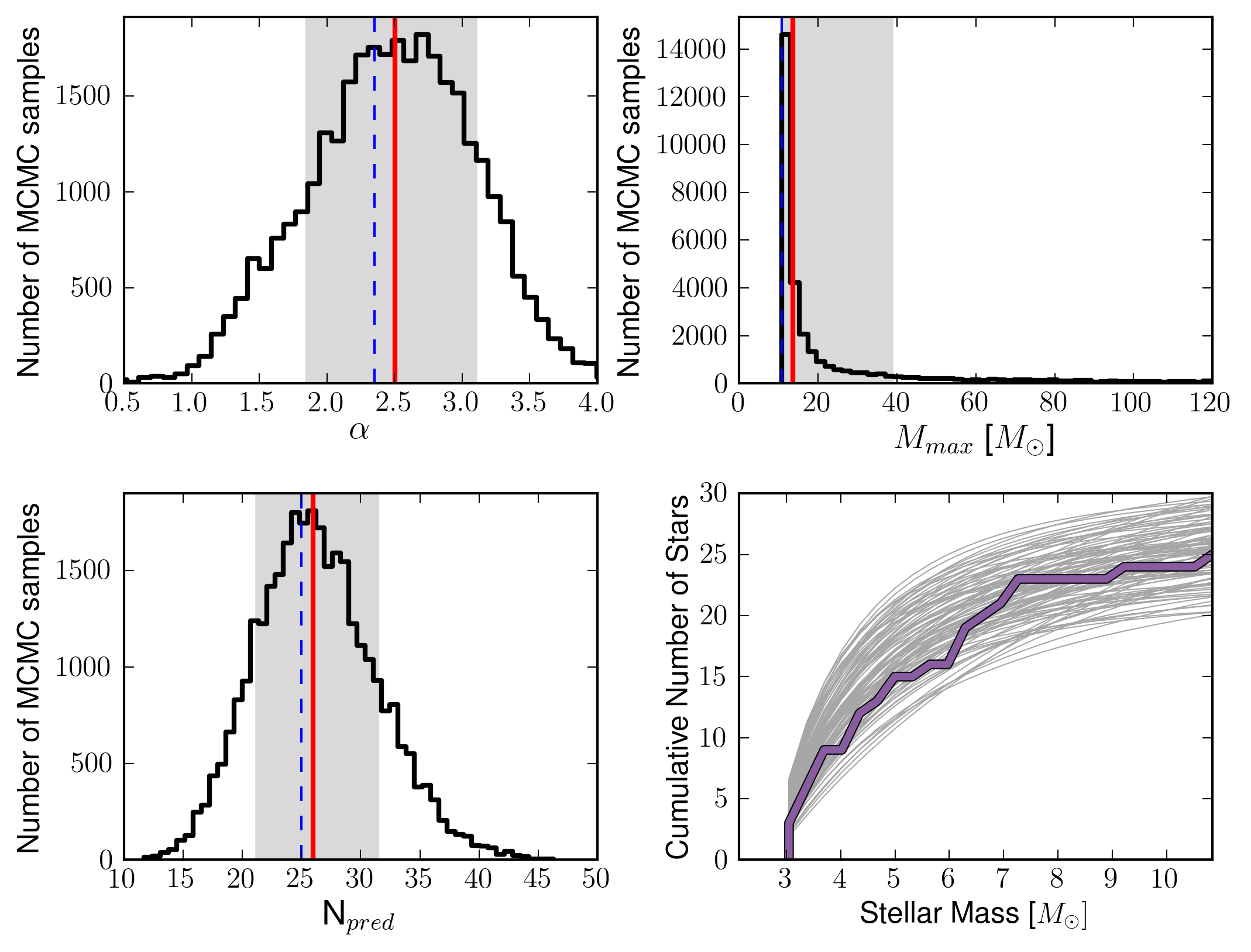}
\caption{The same as Figure \ref{fig:nstar500} only with $N =$ 25 stars and $\Mmaxinput =$ 15 \msun.  Here, we recover $\alpha =$ 2.50$_{-0.65}^{+0.60}$, $\Mmax =$ 13.6$_{-2.3}^{+25.2}$ $\Npred =$ 26$_{-5}^{+5}$.  The recovered fractional precision on $\alpha$ is roughly a factor of 8 larger than for the example considered in Figure \ref{fig:nstar500}.  This exercise demonstrates that it is still possible to place infer the slope of the MF from a sparsely populated cluster.}
\label{fig:nstar25}
\end{center}
\end{figure}

In Figures \ref{fig:nstar500}-\ref{fig:nstar25}, we show select examples for tests of cluster MF recovery in the cases where $N =$ 500, 100, and 25 stars and $\Mmaxinput =$ 60, 30, and 15 \msun.  The first three panels in each figure show the marginalized distributions for $\alpha$, $\Mmax$, and $\Npred$.  For the case of 500 observed stars and $\Mmaxobs =$ 59.9 \msun\ (Figure \ref{fig:nstar500}), we recover $\alpha =$ 2.35$_{-0.07}^{+0.07}$, $\Mmax =$ 63.3$_{-3.1}^{+9.0}$, and $\Npred =$ 500$_{-21}^{+22}$, which are all consistent with their input or observed (in the case of $\Mmax$) values.  
 
Figures \ref{fig:nstar100} and \ref{fig:nstar25} are examples of the same tests applied to populations with fewer stars and smaller dynamic ranges in mass.  In both examples, the recovered parameters are consistent with the input values.  As expected, as the amount of observational information decreases, the constraints become increasingly broad.  However, even in the limit of 25 observed stars and $\Mmaxinput =$ 15 \msun\ (Figure \ref{fig:nstar25}), we find $\alpha =$ 2.50$_{-0.65}^{+0.60}$ and $\Npred =$ 26$_{-5}^{+5}$.  Thus, these ranges contain the input/observed values for each parameter.

In general, these exercises demonstrate probabilistic approach in \S \ref{sec:probimf} can recover MF parameters, even with limited information.  In other words, useful constraints can be derived even for clusters that fail to reach a `gold-standard' level. Importantly, we verify that the recovered MF slope is equivalent to the input IMF slope, as is expected in the case of a single age cluster \citep[e.g.,][]{elm06}.  This finding affirms that this probabilistic technique provides the means to directly infer the slope of the high mass IMF of a given cluster with full and accurate accounting of the associated uncertainties, in the absence of dynamical effects, unresolved binaries, etc., as discussed in \S \ref{sec:caveats}.

\begin{figure*}[]
\plotone{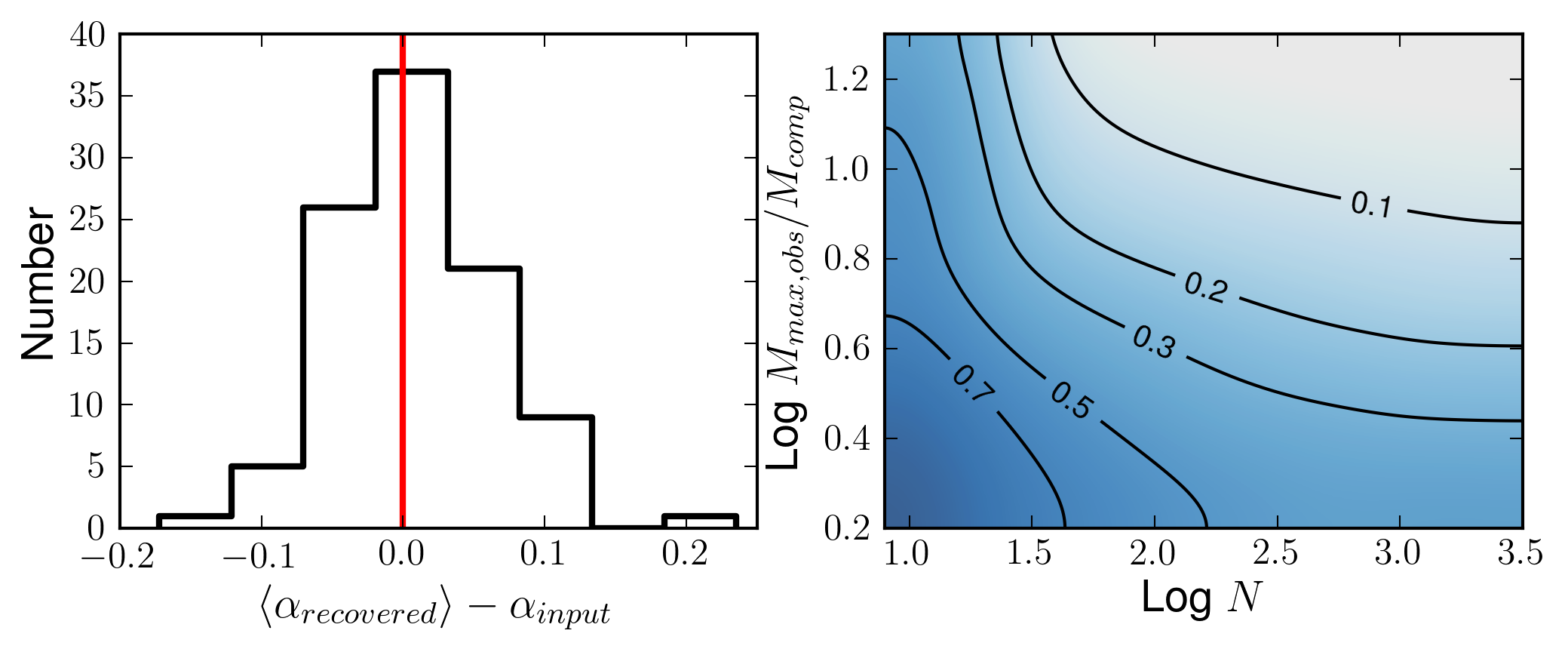}
\caption{The accuracy and precision to which we can recover $\alpha$, in the case of negligible mass uncertainties and a boxcar completeness function.  {\it Left panel} -- The distribution of median values of $\alpha$ from 100 recovered datasets, each with 10$^{4}$ stars and $\Mmaxinput =$ 60 \msun.  The resulting distribution is well-characterized by a normal distribution with $\mu =$ 0.006 and $\sigma =$ 0.07, indicating that this approach to MF recovery is free of systematic biases.  {\it Right panel} -- A 2D map of the theoretical precision to which $\alpha$ can be recovered as a function of the number of stars and dynamic mass range.  This plot indicates that that intrinsic precision (i.e., 0.5$\times$(84th-16th percentiles)) for the recovery of $\alpha$ ranges from $<$ 0.1 in the limit of many stars and a high dynamic mass range to $>$ 0.7 for a poorly populated cluster with a low dynamic mass range. However, under the assumption that masses are independently drawn from the MF, it is possible to combine multiple small clusters to improve precision of MF recover.  The asymmetry in the distribution indicates a slight increase in the recovery of $\alpha$ for a small cluster with a high dynamic mass range relative to a well-populated cluster with a low dynamic mass range.}
\label{fig:alphaprecision}
\end{figure*}

On the other hand, the recovered constraints on $\Mmax$ are less informative about the most massive star that could have formed, i.e., the upper mass limit of the IMF.   We find that the recovered values of $\Mmax$ only constrain $\Mmaxobs$, which is only a lower limit on $\Mmaxinput$, the IMF maximum mass limit for a given cluster.   We therefore suggest that dedicated searches for the most massive star (e.g., spectroscopic surveys) are more suited to constraining the upper stellar mass limit of the IMF than application of this technique \citep[e.g.,][]{mas95, len97, mas03, kud08, cro10, eva11}. Consequently, for the remainder of the paper, we will primarily focus on the recovery of $\alpha$, and only discuss constraints on $\Mmax$ in cases of particular interest.

Extending the above exercises to larger ranges of $N$ and $M_{\max}$ allows us to illustrate the precision and accuracy to which $\alpha$ can be recovered for a diverse set of clusters.  We first check the accuracy of the method to verify that there are no systematic biases in the recovered MF slope. To do this, we simulated 100 different clusters each with 10$^{4}$ stars, to minimize stochastic effects, and $\Mmaxinput =$ 60 \msun, for the same underlying MF.  We recovered the MF for each of the 100 clusters and plot the distribution of the difference between $\alpha_{\mathrm{input}}$ and the median value of $\alpha_{\mathrm{recovered}}$ in the left panel of Figure \ref{fig:alphaprecision}.  The mean of the distribution is 0.006 and 0.5 $\times$ 84th - 16th percentiles, i.e., a Gaussian 1-$\sigma$, is $\pm$0.07.  Based on this set of tests, we see no evidence for systematic biases in the recovery of $\alpha$ in any of the parameter space considered.

Next, we explore the theoretical precision to which $\alpha$ is recovered as a function of $N$ and the dynamic range of the observed masses, as proxied by $\log(\Mmaxobs/\Mcomp)$.  We constructed a grid of model clusters spanning a range in $N$ and $\log(\Mmaxobs/\Mcomp)$ as follows.  We fixed $\Mcomp =$ 3 \msun\ and created 7000 simulated clusters by drawing random values $N$ and $\Mmaxinput$ such that 0.8 $\le$ $\log(N)$ $\le$ 3.5 and 0.2 $\le$ $\log(\Mmaxobs)$ $\le$ 1.3, and discretely sampled the IMFs following the procedure outlined at the beginning of this section.  We recovered the MF slope for each realization and computed $\Delta \alpha =$  0.5 $\times$ (84th - 16th percentile), i.e., a percentile based 1-$\sigma$ equivalent.  To account for fluctuations due to the discrete nature of the sampling, we applied a Gaussian smoothing kernel to the resulting grid with a 1-$\sigma$ width of $\sim$ 5\% of the dynamic range in each variable.  We computed the smoothed and un-smoothed values and found they differ by no more than $\sim$ 7\% at any point.  

We have plotted the results of this exercise in the right panel of Figure \ref{fig:alphaprecision}.  In the limit of high observational information, e.g., $N \sim$ 10$^3$ stars and $\log(\Mmaxobs/\Mcomp)$ $\sim$ 1.1, we see that $\alpha$ is recovered with a precision of $\sim$ $\pm$0.1.  At the other extreme ($N \sim$ 10 stars and $\log(\Mmaxobs/\Mcomp)$ $\sim$ 0.3), the precision in $\alpha_{\mathrm{recovered}}$ decreases significantly to $\gtrsim$ $\pm$ 0.7.  Because the mock data used for these tests are highly idealized, i.e., perfectly known individual stellar masses and completeness,  these values reflect the highest theoretical precision with which $\alpha$ can be inferred from a given number of stars and a measured dynamic mass range.  As we discuss in the next sections, other considerations such as finite mass uncertainties and completeness will degrade these levels of precision, lessening the achievable constraints.

\begin{figure*}[th!]
\begin{center}
\plottwo{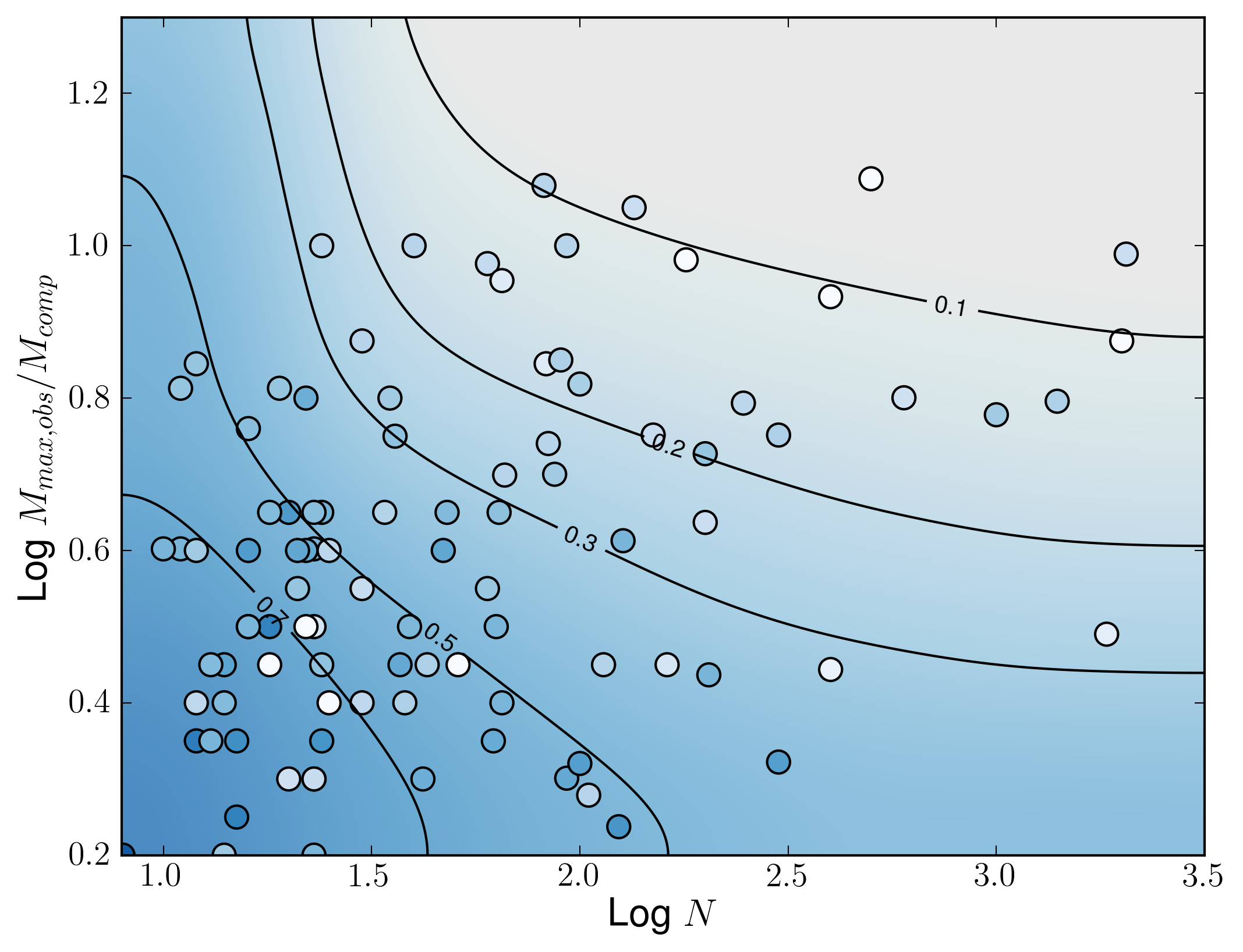}{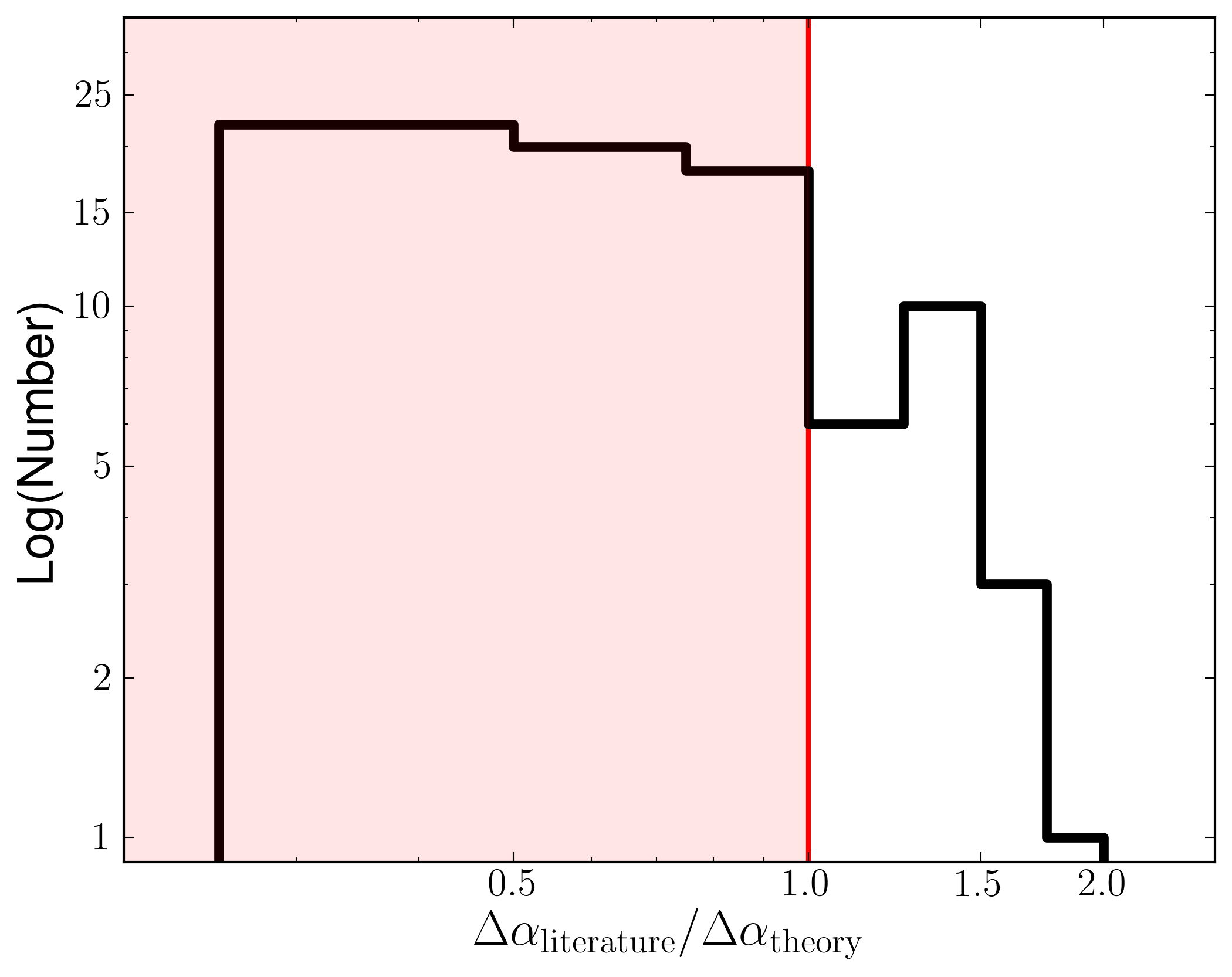}
\caption{\textit{Left --}The 2D map of the theoretical precision to which $\alpha$ can be recovered from Figure \ref{fig:alphaprecision}.   We have over plotted the indicated precision from the literature studies shown in Figure \ref{fig:bastian_fig2} and tabulated in Table \ref{tab1}.   Points that appear darker than the surrounding regions have listed errors above the theoretical precision limits, where as lighter points have errors that are smaller than are theoretically permissible. \textit{Right --} A histogram showing the distribution of the ratio of quoted literature errors and our computed theoretical limits for clusters show in Figure \ref{fig:litcompabs} and tabulated in Table \ref{tab1}.  The red shaded region indicates the region in which literature studies have quoted errors less than the theoretical limit.  The range of ratios varies over an order of magnitude, and approximately $\sim$ 3/4 of the literature studies quote error bars than are below theoretical expectation. }
\label{fig:litcompabs}
\end{center}
\end{figure*}

Additionally, using a least squares fitting routine, we find that $\Delta \alpha$ can be reasonably approximated as a function of $\log(N)$ and $\log(\Mmaxobs/\Mcomp)$ by a 3rd order polynomial

\begin{equation}
\Delta \alpha = \sum_{i=0}^{3} \, \sum_{j=0}^{3} \beta_{i,j} \, (\log(N))^{i} \, (\log(\Mmaxobs/\Mcomp))^{j} \, ,
\label{eq:deltaalpha}
\end{equation}

\noindent where the coefficients $\beta_{i,j}$ are listed in Table \ref{tab2}.  Within the range of $\log(N) =$ 1.5 to 3.0 and $\log(\Mmaxobs/\Mcomp) =$ 0.4 to 1.2, this approximation is generally good to within $\sim$ 20\%.  Outside this range, this analytic expression is noticeably steeper, meaning the extrapolations into regimes of lower and higher information content lead to  over- and underestimates of the true precision.  Nevertheless, when used with a reasonable level of caution, this analytic expression provides a good rule of thumb for the best attainable precision on the MF slope.

Finally, we tested the accuracy and precision to which other $\alpha$ values can be recovered.  Differing values of $\alpha$ dictate the relative distribution of stellar masses, and a significantly steeper MF slope would produce fewer massive stars, potentially resulting in less accurate and/or precise constraints on the MF slope.  To test this potential dependence on $\alpha$, we selected two different, but plausible values of $\alpha$ = 1.8 and 2.8, and repeated all the simulations presented in Figure \ref{fig:alphaprecision}.  We found that both the accuracy and precision to which $\alpha$ can be recovered is statistically consistent in all cases.  There were random variations at the few percent level; however, such small differences are easily accounted for by the discrete nature of the sampling and the finite number of realizations. 

\subsection{Comparing Theoretical Precision Limits with Values from the Literature}
\label{sec:complit}

As an informative exercise, in the left panel of Figure \ref{fig:litcompabs} we have over-plotted the error bars of the literature MF values from Figure \ref{fig:bastian_fig2} on top of our derived theoretically attainable values.   The contrast in colors between the points and shaded contours indicates the level of agreement between the two; points that are lighter than the surrounding region are below the theoretically attainable precision, while darker points are larger than this lower limit.  A cursory inspection of this figure indicates a full range of differences between the two datasets, i.e., there are a number of points with extreme light and dark color contrast.   We have tabulated the literature data used to construct this plot in Table \ref{tab1}.

To better quantify differences in the two datasets, in the right panel of Figure \ref{fig:litcompabs} we have plotted a histogram of the ratios between the theoretical and literature MF precision values.   The histogram shows that the ratios vary over an order of magnitude around unity, with $\sim$ 3/4 of the error bars quoted in the selected literature studies being smaller than the theoretically attainable limit (red-shaded region).  The underestimated error bars are typically lower than the theoretical expectations by a factor of  $\sim$ 2 or more.  Overly small error estimates can give the false impression of a good constraint, which could lead to claims of a significantly unusual MF measurement.  Conversely, large overestimates of the errors on a MF slope indicate that maximal information may not being extracted from the data.  Both cases reinforce the importance of developing a well-defined methodology for correctly characterizing the MF slope of a young cluster.

\begin{figure}[b!]
\begin{center}
\plotone{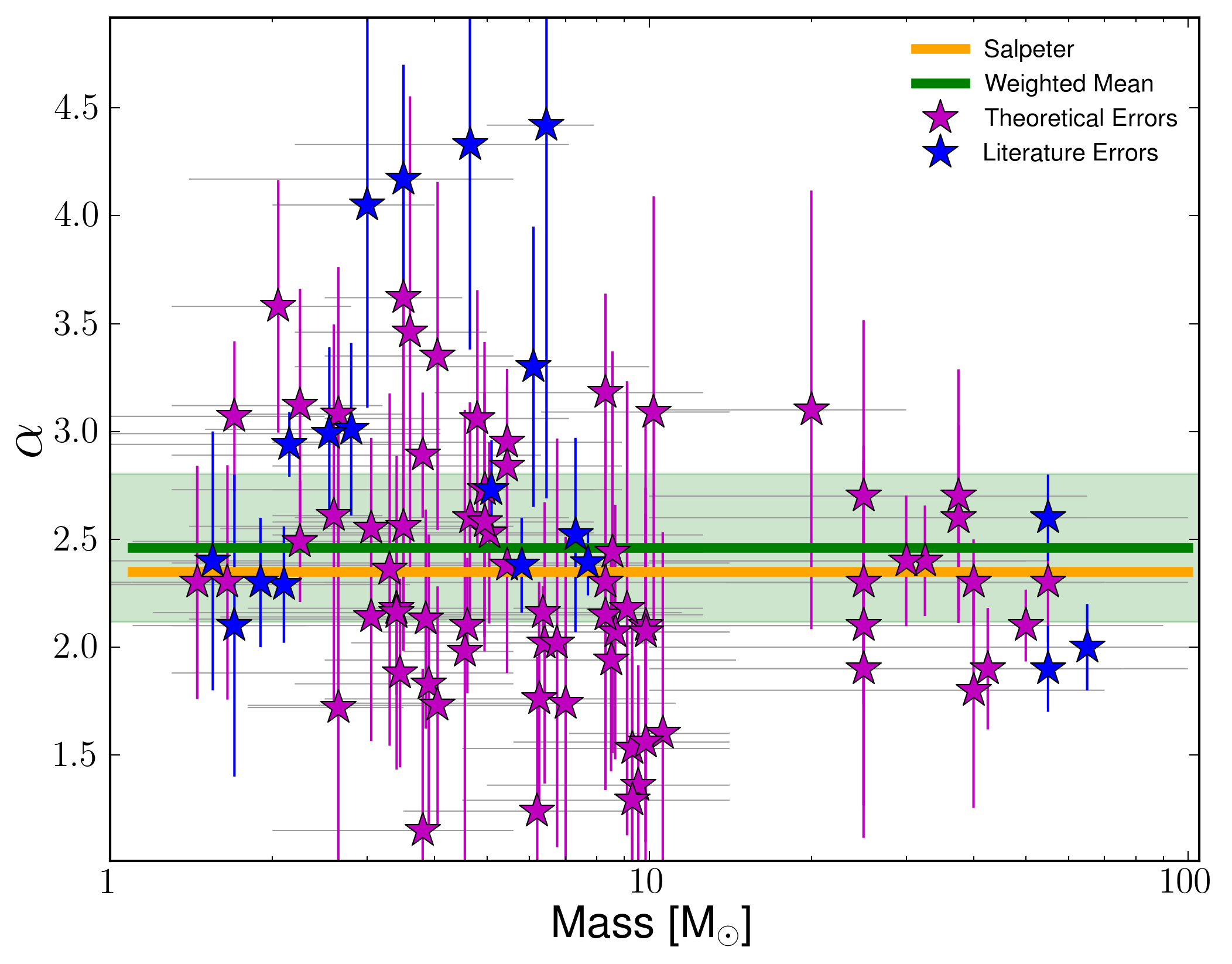}
\caption{A modified version of Figure \ref{fig:bastian_fig2} based on a comparison between the amplitude of the error bars quoted in the literature and the theoretical lower limits derived in this paper.  The blue points are studies whose quoted errors are larger than the theoretical lower limits.  The magenta points are those that had quoted error bars smaller than the theoretical lower limit.  For these points, we have assigned and plotted new error bars equal to the theoretical lower limits.  We then use all the literature data to compute the weighted mean (solid green line) and weighted 1$-\sigma$ (lightly shaded green region) of $\alpha$, which we find to be $\langle \alpha \rangle =$ 2.46 with a 1-$\sigma$ dispersion of 0.35 dex. Note that while we are able to assign more robust error bars to most points, we are unable to assess the accuracy of each value of $\alpha$, which would require re-analysis of each dataset.}
\label{fig:bastian_mod}
\end{center}
\end{figure}

Using the literature measurements of $\alpha$ in tandem with the theoretical lower limit errors, we compute a mean $\alpha$ for the literature values considered. In Figure \ref{fig:bastian_mod}, we have re-created Figure \ref{fig:bastian_fig2}, but have plotted the theoretical lower limits errors for those studies which have underestimated the uncertainties.  These are indicated as magenta points.  Studies whose error bars are larger than the theoretical limit have been left as is, and are plotted as blue points.  From the Figure \ref{fig:bastian_mod} and Table \ref{tab1}, we see there is no discernible trend for which studies, values of $\alpha$, or mass ranges have underestimated error bars.  Using all reported $\alpha$ values, along with the new set of uncertainties, we computed a weighted mean and weighted 1-$\sigma$ uncertainty for this selection of IMF studies and find $\langle \alpha \rangle =$ 2.46 with a 1-$\sigma$ dispersion of 0.35 dex.  This value is consistent with the slope of a Salpeter and Kroupa IMFs, $\alpha$ = 2.35 and 2.3, respectively. However, the scatter is large ensuring the mean value is also consistent with the Kennicutt IMF \citep[][]{ken83}, $\alpha =$ 2.5, and the Scalo IMF \citep[][]{sca86}, $\alpha \sim$ 2.6.  Although our computed mean value of $\alpha$ depends on our selection of literature studies and the accuracy of each study, the broad scatter reinforces the notion that our knowledge of the high mass IMF slope is far from secure.

\begin{figure}[b!]
\begin{center}
\plotone{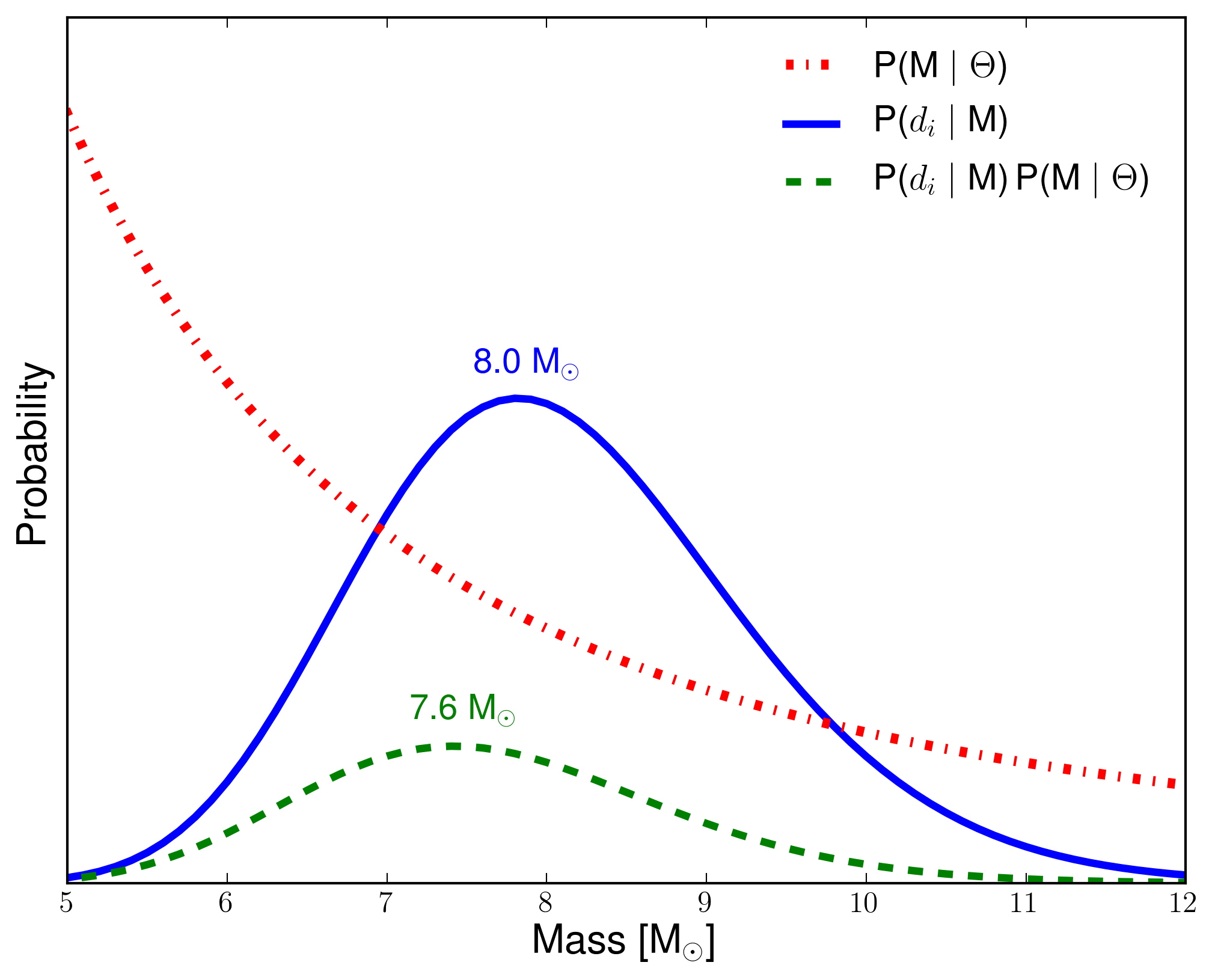}
\caption{A schematic demonstration of the importance of considering the MF as a prior in the determination of individual stellar masses.  The result of photometric SED fitting for a stellar masses, i.e., $p(\vec{d} \midline M)$, is denoted by the blue line.  The MF with $\alpha =$ 2.35 (red dot-dashed line), indicates that lower stellar masses are more likely than higher stellar masses.  Therefore, the best mass estimate for the star comes from the convolution of these two functions (green dashed line).  Not taking the MF into account when deriving stellar mass PDFs can lead to noticeable overestimates of the star's true mass.}
\label{fig:mass_schematic}
\end{center}
\end{figure}

\section{Incorporating Uncertainties in Stellar Masses}
\label{sec:massunc}

Up to this point, we have explored placing constraints on MF parameters presuming data with negligible individual mass uncertainties.  However, in practice, the mass of a star is not perfectly known. Degeneracies with extinction, metallicity, etc., along with uncertainties in stellar models all contribute to uncertainties in estimating the mass of a star.   
 
Clearly, the uncertainty on the mass of an a star needs to be incorporated into the pPDF for realistic MF determination. To understand the relationship between MF determination and stellar mass uncertainties, it is useful to review how stellar masses are measured.  

Most individual star masses are inferred from  photometric spectral energy distribution (SED) fitting, i.e., by comparing photometric observations of a star,  $\vec{d}$, with predicted fluxes from stellar models.  Specifically, one can construct a set of predicted fluxes in the observed bands by for various combinations of log($T_{eff}$), log($g$), $Z$, $A_{V}$, $M$, age, etc., and then perform a comparison, e.g., $\chi^2$-minimization, between the observed and predicted fluxes.  From this process one is left with a multi-dimensional distribution for the probability of the star's parameters given the data, i.e., $p( \vec{d_i} \midline$ log(T$_{eff}$), log($g$), $Z$, $A_{V}$, $M$, age).  One can then compute the likelihood of the data given a mass, i.e., $p(\vec{d} \midline M)$, by marginalizing over all other model parameters.  The resulting PDF for a star's mass accounts for correlations and uncertainties in the all other parameters.

While the quantity $p(\vec{d} \midline M)$ is often thought to reflect the probability of a star's mass, the desired quantity is in fact, $p(M \midline \vec{d})$, the probability that a star has a particular mass given the observations.  To determine this quantity, one can apply Bayes's theorem, which yields

\begin{equation}
p(M \midline \vec{d}) = \frac{1}{\zeta} \, p(\vec{d} \midline M) \, p_{\mathrm{prior}}(M) \; ,
\label{eq:bayesmass}
\end{equation}

\noindent where $p_{\mathrm{prior}}(M)$ reflects any prior knowledge about the probability of individual stellar masses, and $\zeta$ is a normalization factor needed to make $p(M \midline \vec{d})$ a true probability. 

A temptingly simple assumption for $p_{\mathrm{prior}}(M)$ would be that all stellar masses are equally likely, i.e.,  $p_{\mathrm{prior}}(M) =$ constant.  In this case, ${p}(M \midline \vec{d})$ and $p(\vec{d} \midline M)$ can be treated interchangeably, which is at least implicitly assumed in many published analyses \citep[e.g.,][]{mas95, mas95b, pag00, bra08, bia12}.

However, it is clearly more appropriate to adopt the cluster's MF as a prior on the distribution of individual stellar masses.  As illustrated in Figure \ref{fig:mass_schematic}, failure to consider the effects of the MF when determining the mass of a star will lead to a biased mass relative to the star's true mass in the presence of significant mass errors.

Adopting the MF from Equation \ref{eqnarray:MFobs} as a prior on stellar mass, we can now write down an expression for the probability of the data, $\vec{d}$, given MF parameter, $\vtheta$, as:

\begin{align}
p(\vec{d} \midline \vtheta, \mathrm{obs}) &= \int p(\vec{d}\midline M)  \, \pMFobs \, {\mathrm d}M \nonumber \\
&= \frac{p(\mathrm{obs} | \vec{d}) \int p(\vec{d_{i}}|M) \,  \pMFup \, {\mathrm d}M}{p(\mathrm{obs} \midline \vtheta)} \; ,
\label{eq:witherrors}
\end{align}

and

\begin{equation}
p(\mathrm{obs} \midline \vtheta) = \int p(\mathrm{obs} \midline \vec{d}) \, p(\vec{d} \midline \vtheta) \, {\mathrm d}\vec{d} \; ,
\label{eq:witherrors_norm}
\end{equation}

\noindent where $p(\vec{d} \midline M)$ is the likelihood of the observed data given a true mass, as determined by photometric SED fitting, and $p(\mathrm{obs} \midline \vtheta)$ is the normalization factor, which represents the probability of observing the data (i.e., fluxes) given the MF model parameters $\vtheta$.

Using Bayes's theorem, we can write the probability for the parameters, $\vtheta$ and $\Npred$, given the observed data, $\vd$ as:

\footnotesize
\begin{align}
&p_{\mathrm{post}}(\vtheta, \Npred \midline \vd, \mathrm{obs})  =  p_{poi}(N \midline \Npred) \times \nonumber \\ &\sum_{i=1}^{N} \frac{\pobsd  \, \int p(\vec{d_i}|M) \, \pMFup \, {\mathrm d}M}{p(\mathrm{obs} \midline \vtheta)} \,  p_{\mathrm{prior}}(\vtheta) \,  p_{\mathrm{prior}}(\Npred) \; , 
\label{eq:longposterior1}
\end{align}
\normalsize

\noindent which is the pPDF in the case of non-negligible mass uncertainties. 

\subsection{Illustrating the Effects of Log-Normal Mass Uncertainties}
\label{sec:loggaussmass}

As an example of how to fold in finite mass uncertainties in practice, we consider the case that each star has an observed mass PDF that can be represented by a log-normal function.  The assumption of a log-normal not only provides an analytically tractable solution, which aides in the clarity of this paper, but also log-normal approximations (i.e., $\log(M)\pm \delta m$) have been used in the literature to summarize the inferred masses of stars \citep[e.g.,][]{mas98, mas11, bia12}.

Consider a set of $N$ stars each of which has an associated mass PDF. We believe that each observed mass PDF is the convolution of a single true mass, $M_i$, with a log-normal noise model (due to observational uncertainties, the flux-to-mass conversion process, etc.).  The inferred mass for each star can then be summarized as having a mean value of $\Mbar_i$ and a fractional mass uncertainty, $f_i \equiv M_i/ \Mbar_i$, i.e., the linear width of the mass PDF scales as $\sim$ $e^f$.  Considering only the upper portion of the IMF, the stars we observe can have true masses ranging from $\Mmin$ to $\Mmax$, while the inferred masses will range according to the stellar evolution models and the observed completeness function, and thus have limits of $\Mcomp$ and $\Mlim$.  For simplicity, we will adopt a boxcar completeness function written in terms of $\Mbar$,

\begin{equation}
\pobsMbar = 
\begin{cases}
1, & \Mcomp \le \Mbar  \le \Mlim\ M_{\odot}  \\
0, & \text{otherwise \; .}
\end{cases}
\label{eq:masscompleteness}
\end{equation}

Having defined the relevant variables and the ranges, we can return to the derivation. The goal of this process is to place constraints on the MF parameters, given the set of log-normal PDFs for the inferred masses, and Equation \ref{eq:witherrors} provides the framework to complete this task.  Substituting a log-normal mass model for $p(\vec{d_i}\midline M)$ and making the coordinate transformation $X\equiv \lnn M/\overline{M}$, allows us to write:

\begin{align}
&p(\vec{d_i} \midline \vtheta, \mathrm{obs}) = \nonumber \\ &c_{\mathrm{MF_o}}(\vtheta)\, \overline{M}^{-\alpha} \, \frac{\pobsMbar}{\sqrt{2\pi}f}  \int _{\ln\left(\frac{\Mmin}{ \overline{M}} \right)}^{\ln\left(\frac{\Mmax}{ \overline{M}} \right)}  e^{ -\frac{X^2}{2f^{2}} } \, e^{(1-\alpha) X} {\mathrm d}X \; ,
\label{eq:witherrors2}
\end{align} 

\noindent where the completeness is over the observed masses, $\pobsMbar$, the integration is over the permissible range of true masses, and we have made the simplifying assumption of a constant fractional mass uncertainty for all masses, i.e., $f_i \rightarrow f$.  This assumption has been made for clarity in this derivation, but in practice, $f_i$ is not likely to be constant.  Stars with $M$ $\gtrsim$ 25 \msun\ are less constrained that those of lower mass stars \citep[e.g.,][]{mas95}, owing to both uncertainties in massive star evolution models, as well as degeneracies in the optical and near-UV photometric colors of massive stars, which make it difficult to precisely characterize extremely massive stars.

Integration of Equation \ref{eq:witherrors2} yields a closed form solution:

\small
\begin{align}
&p(\vec{d} \midline \vtheta, \mathrm{obs}) =  \nonumber \\ & c_{D}(\vtheta) \, c_{MF_o}(\vtheta) \,  \overline{M}^{-\alpha} \, \pobsMbar \, e^\frac{(\alpha - 1)^2 f^2}{2} \times \nonumber \\ & \left(\eta \left[\frac{(\alpha-1)f^2
 + \ln(\frac{\Mmax}{\overline{M}})}{f}\right] - \eta \left[\frac{(\alpha-1)f^2 + \ln(\frac{\Mmin}{\overline{M}})}{f}\right] \right) \; ,
\label{eq:witherrors4}
\end{align}
\normalsize

\noindent where

\small
\begin{align}
&c_{\mathrm{D}}(\vtheta)^{-1} =  \nonumber \\ &c_{MF_o}(\vtheta) \, e^\frac{(\alpha - 1)^2 f^2}{2} \int_{\Mcomp}^{\Mlim} \overline{M}^{-\alpha} \,  \pobsMbar \times \nonumber \\ &  \left(\eta \left[\frac{(\alpha-1)f^2 + \ln(\frac{\Mmax}{\overline{M}})}{f}\right] - \eta \left[\frac{(\alpha-1)f^2 + \ln(\frac{\Mmin}{\overline{M}})}{f}\right] \right) {\mathrm d}\overline{M} \; , 
\label{eq:witherrors3}
\end{align}
\normalsize

\noindent and $\eta$ is the normalized cumulative distribution function of a standard Gaussian PDF with $\mu =$ 0 and $\sigma =$ 1:

\begin{equation}
\eta(X) = \frac{1}{\sqrt{2\pi}} \int_{-\infty}^{X} e^{-\frac{x^2}{2}} dx \; .
\end{equation}

In the limit of very accurate mass estimates, i.e., $f\rightarrow 0$, the expression for ${p}(\vec{d_i}|\vtheta)$ reduces to a scenario where the masses are perfectly known, as in Equation \ref{eq:probimf}. 

Having derived an expression for $p(\vec{d_i} \midline \vtheta)$, we can now write the pPDF for the MF parameter as:
 
\begin{align}
&p_{\mathrm{post}}(\vtheta, \Npred \midline \{\vec{d}_i\}, N, \mathrm{obs})  = \nonumber \\  &p_{\mathrm{poi}}(N | \Npred) \, \sum_{i=1}^{N}\lnn p(\vec{d_i}|\vtheta, \mathrm{obs}) \,  p_{\mathrm{prior}}(\vtheta)  \, p_{\mathrm{prior}}(\Npred) \; ,
\label{eq:longposteriormass}
\end{align}

\subsection{Illustrative Examples with Mass Uncertainties}
\label{massexamples}

In this section, we present several examples of MF recovery in the presence of non-negligible mass uncertainties.  However, before delving into specific cases, it is instructive to first examine how log-normal mass PDFs affect the observed MF.

\subsubsection{Schematic Examples with Continuous Mass Functions}
\label{sec:continuousunc}

\begin{figure}[t!]
\plotone{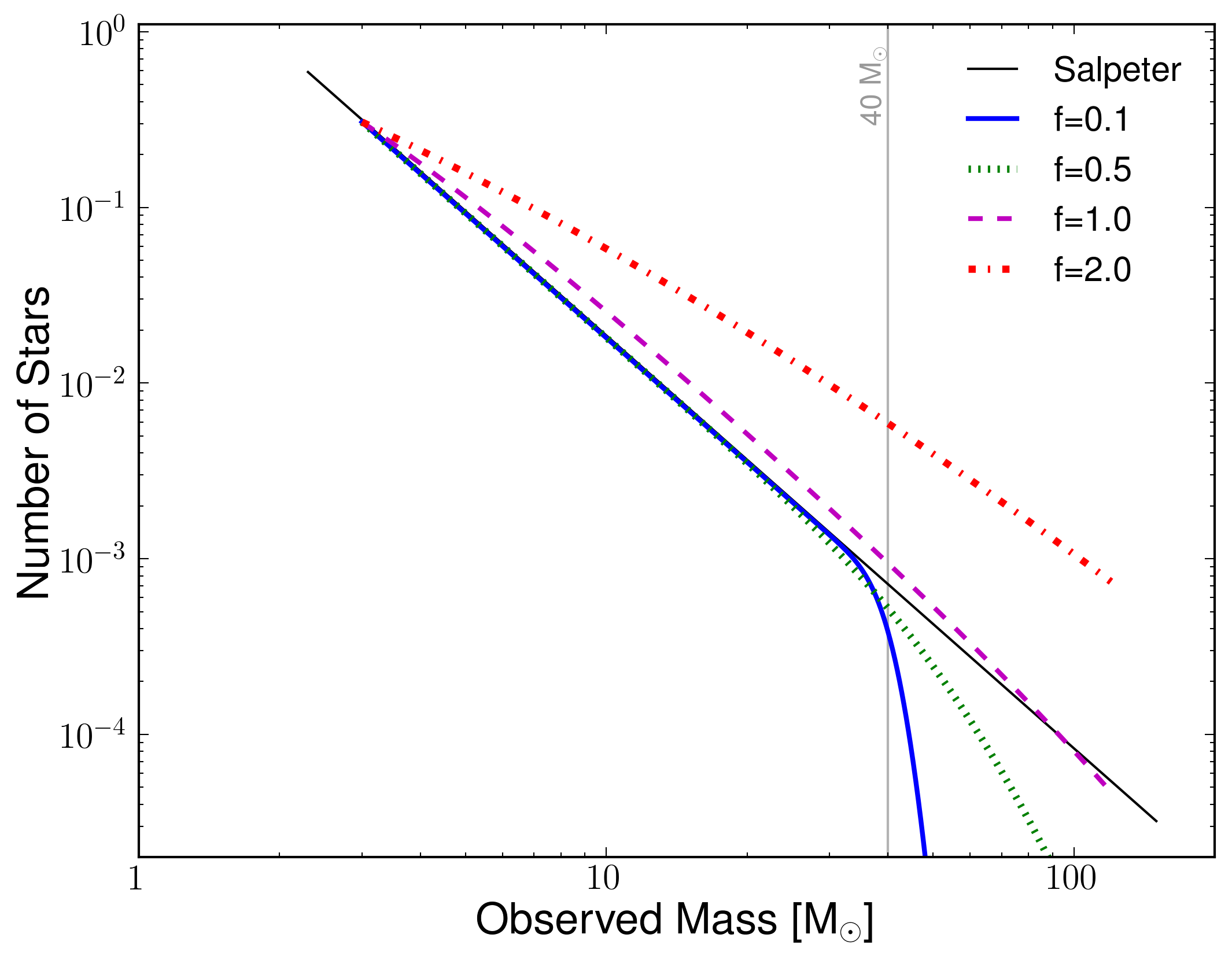}
\caption{A schematic demonstrating the influence of log-normal individual stellar mass PDFs on the MF.  Each distribution was constructed assuming $\Mmaxinput =$ 40 \msun\ (grey line) and $\alpha_{input} =$ 2.35 (black line).  We then varied the value the fractional mass uncertainties, $f$, as indicated. Small fractional mass uncertainties lead to a slightly steeper than Salpeter MF slope, and predicted the presence of masses larger than $\Mmaxinput$.  For extreme values of $f$, e.g., 1, the observed MF slope is flatter than Salpeter, with observed values of $\Mmax$ significantly larger than $\Mmaxinput$.  Failure to properly account for mass uncertainties when modeling the observed MF can lead to biases in the recovery of $\alpha$ and $\Mmax$. See \S \ref{sec:loggaussmass} for more discussion.}
\label{fig:masspdf_schematic}
\end{figure}

In Figure \ref{fig:masspdf_schematic}, we illustrate the changes in the (continuous) observed MF for different values of $f$. For reference, the solid black line represents a Salpeter IMF from 3 to 120 \msun.  The other plotted lines represent MFs generated using Equation \ref{eq:witherrors2} for select values of $f$ (0.1, 0.5, 1.0, 2.0).  In each case,  $\Mmaxinput = $ 40 \msun\ and each MF was normalized to have the same number of 3 \msun\ stars as a Salpeter MF.

For the smallest fractional mass uncertainty, $f =$ 0.1, we see that the observed MF (solid blue line) is identical to Salpeter until $\sim$ 37 \msun.  At this point it begins to deviate from the Salpeter MF, and predicts the presence of stars above the nominal value of $\Mmaxinput =$ 40 \msun. Taken at face value, this MF appears slightly steeper and it implies a finite probability that $\Mmaxobs$ is larger than $\Mmaxinput$. The MF for $f =$ 0.5 exhibits the same qualitative behavior, with an even steeper slope and a significantly larger apparent value of $\Mmaxobs$.  Naive fitting of the observed MF, without taking mass errors into account, would therefore lead to biased measurements of both $\alpha$ and $\Mmax$.

For the extreme values of $f$=1 and 2, the observed MFs flatten out relative to Salpeter, and each predicts a significant population of stars with masses in excess of 100 \msun.  For such extreme uncertainties, a significant number of stars with true masses below 3 \msun\ will have inferred masses above the completeness limit.

This schematic also illustrates the need to alter the prior restrictions placed on $M_{\max}$.  In the case of no mass uncertainties, the inferred and true masses are identical, implying that $M_{\max} \ge \Mmaxobs$ was a reasonable requirement.  However, for finite mass uncertainties, $\Mmaxobs$ is likely an overestimate of $\Mmaxinput$, i.e., the maximal inferred mass is larger than the maximal true mass, invalidating the previous assumption.  For the exercises involving finite mass uncertainties, we therefore only require that $M_{\max} \le \Mlim$, the larger mass permitted by stellar evolution models.

\subsubsection{Illustrative Examples with Simulated Clusters}
\label{sec:discreteunc}

We now use mock data to explore recovery of the MF in the case that each mass has a log-normal PDF.  To generate the simulated data, we construct a convolution between the MF a log-normal noise model with $\alpha =$ 2.35 (Equation \ref{eq:witherrors2}).  We then draw 10$^6$ stars from this function, apply the boxcar completeness function from Equation \ref{eq:masscompleteness}, and randomly subsample the mass list to obtain the desired number of stars.   

\begin{figure}[b!]
\begin{center}
\plotone{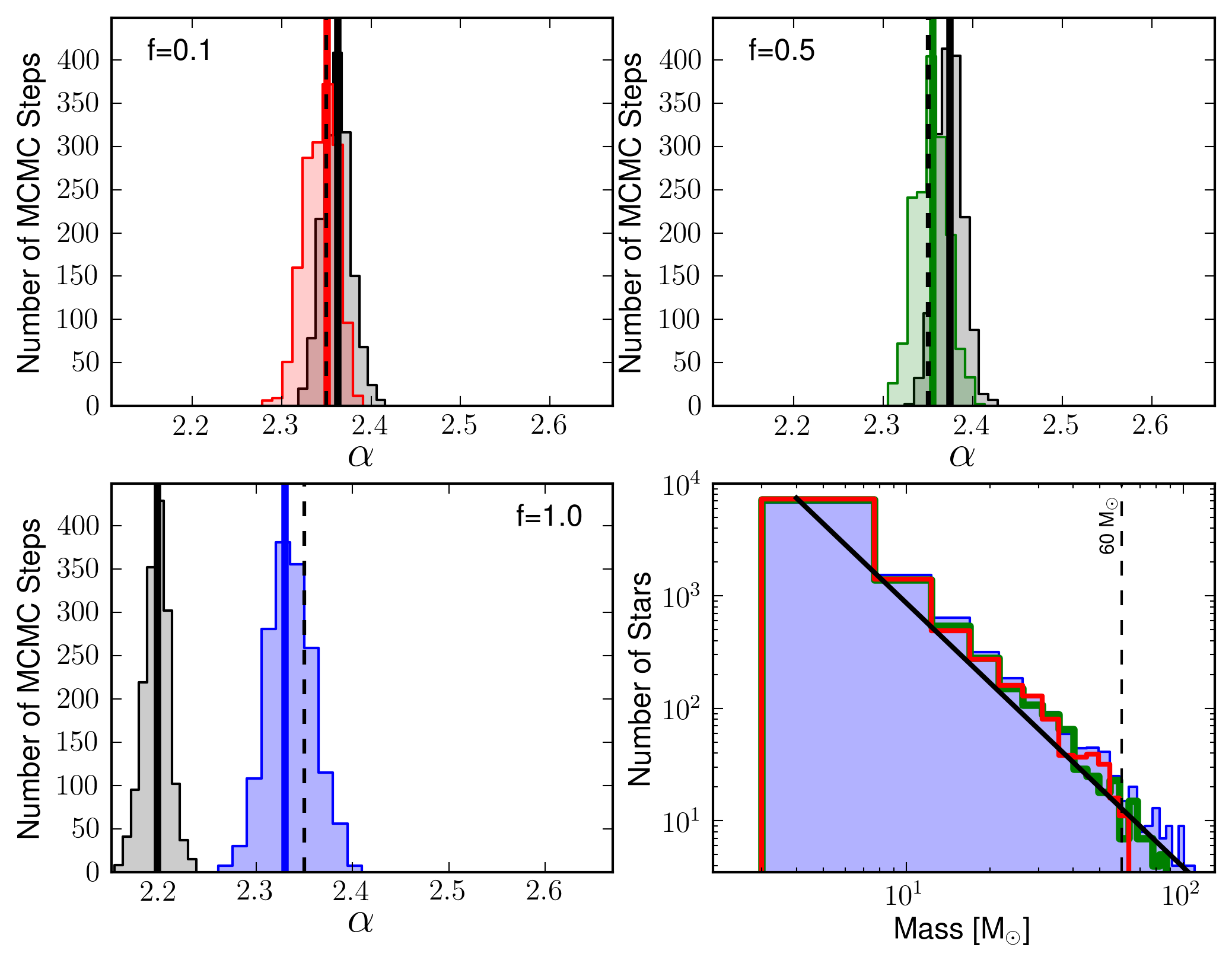}
\caption{The recovered marginalized distributions for $\alpha$ when accounting (colored histograms) and not accounting (grey histograms) for individual stellar mass uncertainties in the MF model (colored histograms) and not including mass uncertainties (grey histograms).  Each star has a log-normal mass PDF with a fractional uncertainty, $f \equiv M/ \Mbar$, whose assumed constant value is indicated.  For each value of $f$ the noisy (i.e., accounting for mass uncertainties) and noiseless (i.e., not accounting for mass uncertainties) models were applied to the same data, which always had $\Mmaxinput =$ 60 \msun and $N =$ 10$^{4}$ stars.  In each of the first three panels, the true MF slope (Salpeter), is indicate by the dashed black line.  Application of the noisy model yields recovery of the colored histograms, whose median value is indicated by the solid colored line.  The grey histograms are the result of applying the noiseless models, and the solid black line represents the median of this distribution.   For reasonable fractional mass uncertainties, e.g.,  $f \lesssim$ 0.5, there are only small differences in the MF slopes recovered by the two methods.  For the extreme case, $f =$ 1.0, modeling the mass uncertainties yields near perfect recovery, while failure to do so results in a systematically flatter MF slope.  In the lower right panel, we show the MF distributions used in each case, and have included a Salpeter MF (with an arbitrary vertical scaling applied) for reference.}
\label{fig:massunc60}
\end{center}
\end{figure}

\begin{figure}[t!]
\begin{center}
\plotone{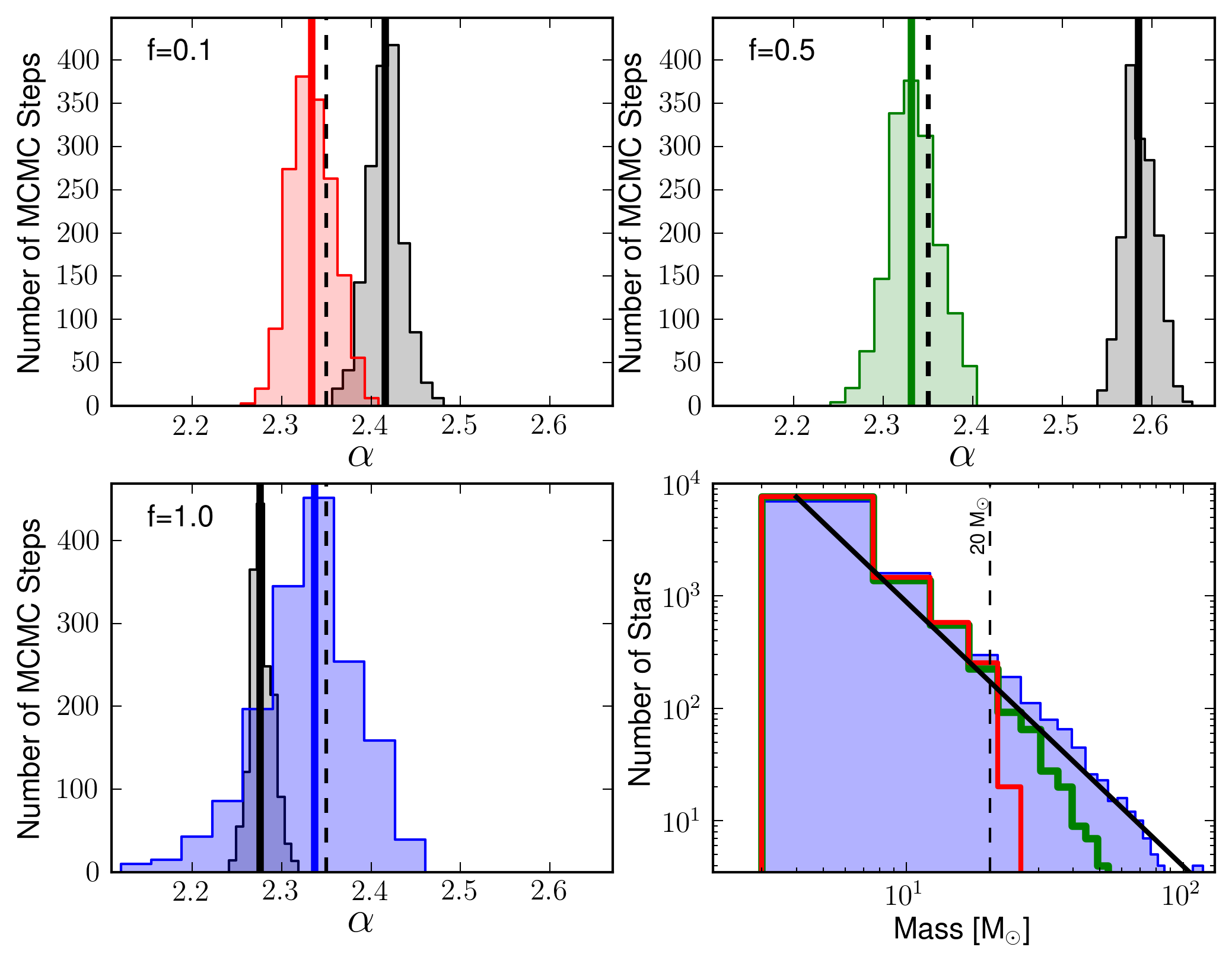}
\caption{The same as Figure \ref{fig:massunc60} only with $\Mmaxinput = $ 20 \msun.  In this case, the differences in the recovered MF slopes between the two methods are more pronounced, even for modest values of $f$.  These examples demonstrate the importance of including a good characterization of mass uncertainties in order to measure the MF slope.}
\label{fig:massunc20}
\end{center}
\end{figure}

In Figures \ref{fig:massunc60} and \ref{fig:massunc20}, we show results from the recovery of $\alpha$ for select simulated clusters.  In each case, the mock data consisted of 10$^{4}$ stars (to minimize random noise effects) and discrete values of $f =$ 0.1, 0.5, and 1.0.  To illustrate the effects of the observed dynamic mass range, we also considered two different values of $\Mmaxinput$ = 20 \msun\ and 60 \msun.  As a baseline for comparison, we applied two versions of the code to each dataset, one that included modeling of the mass uncertainties (the `noisy' MF model; Equation \ref{eq:longposteriormass}) and the simple model MF (the `noiseless' MF model; Equation \ref{eq:longposterior}).  In each panel, the dashed line indicates the input value of  $\alpha$ while the solid line indicates the median value for each marginalized PDF.

In Figure \ref{fig:massunc60}, we show the recovery of $\alpha$ for  $\Mmaxinput =$ 60 \msun.  In the case of small mass errors ($f$ = 0.1), both the noisy (red) and noiseless (grey) model MF models provide excellent recovery of  $\alpha_{\mathrm{input}}$.  The small level of offset between the results are consistent with the expected scatter ($\sim$ $\pm$ 0.03) in the case of $N =$ 10$^{4}$ stars.  The two methods also return the same level of precision on $\alpha_{\mathrm{recovered}}$, as indicated by the comparable widths of the two distributions.

For significantly larger mass errors ($f =$ 0.5), both the noisy (green) and noiseless (grey) models still provide excellent recovery of $\alpha$.  The distributions again share a comparable width.  Although the noiseless PDF appears shifted toward slightly higher values of $\alpha$, the median recovered value is within the expected scatter for a cluster with $N =$ 10$^{4}$ stars.

Extending this exercise to extremely large mass errors ($f =$ 1.0) reveals significant discrepancies between results from the noisy (blue) and noiseless (grey) models.  Based on the schematic in Figure \ref{fig:masspdf_schematic}, the observed MF should be flatter than Salpeter, which is what is the noiseless model returns, with the median $\alpha_{\mathrm{recovered}} =$ 2.20.  Although the distribution is comparable in width to those in the other panels of the plot, it does not overlap the true value of $\alpha$. In contrast, application of the noisy model recovers $\alpha$ to excellent accuracy, although to a lower degree of precision. 

To test the sensitivity of MF slope recovery to $\Mmax$, we conducted the same exercise, only with $\Mmaxinput =$ 20 \msun.  The recovered distributions for $\alpha$ are shown in Figure \ref{fig:massunc20}.  For the lower value of $\Mmaxinput$, we see more pronounced differences in the results from the noisy and noiseless models.  For both $f =$ 0.1 and 0.5, the noiseless model recovers systematically higher values of $\alpha$ relative to the input.  While only a modest bias is present in the case of $f =$ 0.1, the offset is quite substantial when $f =$ 0.5.  In contrast, the recovered values of $\alpha$ from application of the noisy MF model are in excellent agreement with $\alpha_{\mathrm{input}}$.  Once again, the widths of the noisy model distributions are slightly broader than the noiseless model results.  However, the slight decrease in precision is a small tradeoff relative to inferring a significantly more accurate MF slope.  

In the case of $f =$ 1.0, application of the noiseless MF model to the data yields an only slightly biased ($\lesssim$ 0.1) median value of $\alpha$.  This is perhaps somewhat surprising, given the extremely large fractional uncertainties.  However, this small amount of disparity is purely coincidental.  The combination of $f$ and $\Mmaxinput$ conspired to produce an observed MF slope that appears close to the true MF slope.  Such a case might lead to the conclusion of a precise constraint on a MF that is statistically different from Salpeter, despite the large mass errors; however, such a conclusion would clearly be false.  In contrast, modeling the mass uncertainties results in an accurate recovery of the true MF, albeit with decreased precision, as previously discussed.

Overall, we find that incorporating the effects of mass uncertainties into MF modeling results in an unbiased recovery of the MF.   In contrast, failure to model even moderately large mass uncertainties, e.g., $f \gtrsim$ 0.5,  can lead to biases that are difficult to quantify, which can undermine a physically meaningful interpretation of the results.    Additionally, neglecting to model mass uncertainties can lead to a severely low estimate of $\Mmax$, which is equivalent of underestimating the cluster's age.  Finally, as previously mentioned, in practice $f_i$ is unlikely to be constant, as the inferred properties of stars with $M$ $\gtrsim$ 25 \msun\ are far more uncertain than lower mass stars \citep[e.g.,][]{mas95}.  Including a non-constant value of $f$ is straightforward in the presented framework. 

\section{Incorporating the Effects of Observational Completeness}
\label{sec:completeness}

\subsection{Preface}
\label{sec:completenesspreface}

We now consider the effects of observational completeness on the inference of MF parameters.  For clarity in the mathematics and the examples, we return to the case of perfectly known masses.  In practice, both completeness and mass uncertainties need to be simultaneously modeled, but they typically affect opposite ends of the observed mass spectrum, meaning that they are largely not degenerate in terms of the data they impact.  

When deriving the general pPDF for a cluster's MF (Equation \ref{eq:posterior}), we implicitly included the completeness function in the probability of a star being on our data list, $\pMFobs$.  However, writing the pPDF to explicitly include the effects of completeness provides a better illustration of the role it plays.  Using  Equation \ref{eqnarray:MFobs}, we can write the pPDF for a general completeness function as:

\footnotesize
\begin{align}
&{p}_{\mathrm{post}}(\vtheta, \Npred \midline \{M_{i}\}, N, \mathrm{obs}) \nonumber  =  p_{\mathrm{poi}} (N \midline \Npred) \times \\ & \sum_{i=1}^{N} \bigl (\pobsM \, c_{MF_o}(\vtheta) \, M_i^{-\alpha}\bigr ) \,  p_{\mathrm{prior}}(\vtheta) \,  p_{\mathrm{prior}}(\Npred) \, ,
\label{eq:comppdf}
\end{align}
\normalsize

\noindent Here we see that the completeness enters into the pPDF as a linear weight on the MF.  In the limit that the $\pobsM =$ constant, it will have no effect on the pPDF as all points are equally weighted.  In the next section we illustrate how a more realistic completeness function affects the recovery of the MF.

\subsection{Illustrating the Effects of a Linear Completeness Function}
\label{sec:linearcompleteness}

In practice, a boxcar function is not a realistic representation of observational completeness.  Completeness functions in clusters vary widely in form depending largely on the surface brightness of the cluster and the resolution of the observations.  In this section, we consider a linear ramp completeness function.  Such a linear completeness function is analytically tractable and provides a reasonable first order approximation to more realistic completeness functions presented in the literature \citep[e.g.,][]{and08, gen11}.  The linear ramp function, which ranges from 0 to 100\% completeness, can be written piecewise as

\begin{equation}
\pobsM = 
\begin{cases}
1,  \, & M > M_{\mathrm{Cmax}} \\
0, \, & M < M_{\mathrm{Cmin}}  \\
\frac{M-M_{\mathrm{Cmin}}}{M_{\mathrm{Cmax}} - M_{\mathrm{Cmin}}}, & \, M_{\mathrm{Cmin}} \le M \le  M_{\mathrm{Cmax}} \; , \\
\end{cases}
\label{eq:truecompleteness}
\end{equation}

\noindent where $M_{\mathrm{Cmax}}$ and $M_{\mathrm{Cmin}}$ are the upper and lower bounds of the linear portion of the completeness function.   

We again turn to mock data with no mass uncertainties to illustrate MF recovery in the case of a linear completeness function.  Specifically, we consider the cases of how perfectly known, slightly incorrect, and extremely incorrect knowledge of the completeness function affects MF recovery.

To generate the mock data we followed the general description in \S \ref{perfectdata}, with a few modifications.  After drawing a list of masses from the power-law MF, we applied the `true' completeness function to the mass list.  Motivated by data from PHAT, for this exercise, the true completeness function is always a linear ramp between $M_{\mathrm{Cmin}} =$ 2 and  $M_{\mathrm{Cmax}} =$ 4 \msun, with a completeness fraction of 0 for stars with $M$ $\le$ 2 \msun\ and 1 for  $M$ $\ge$ 4 \msun.  After applying this completeness function to the mass list, we then rejected stars from the list with a probability equal to their completeness fraction, e.g., a mass with $\pobsM  =$ 0.3 has a 30\% chance of being observed.  We then perform a uniform draw from the remaining stars to arrive at the desired number of stars.   This set of masses constitutes the true mass list.  To examine effects of a non-perfectly known completeness function, we followed the same procedure as descried above, only the values of $M_{\mathrm{Cmin}}$  and $M_{\mathrm{Cmax}}$ are different when applied to the mock data and when used to model the completeness in the MF recovery.

\begin{figure}[t!]
\begin{center}
\plotone{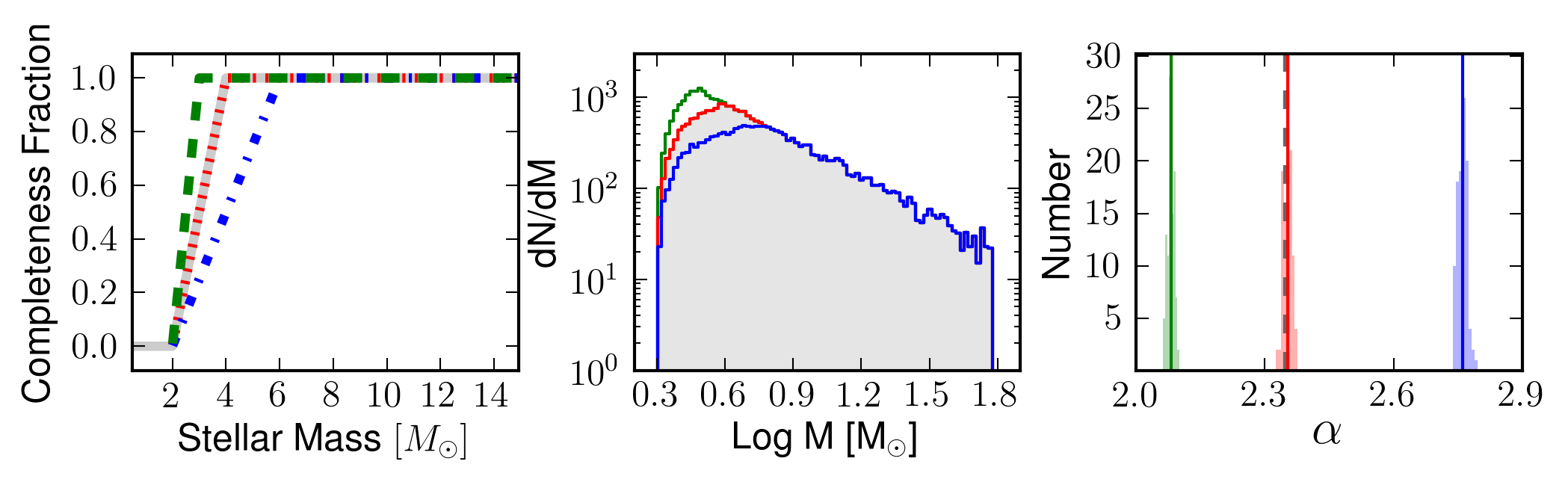}
\caption{The recovery of $\alpha$ from from various linear completeness functions.  {\it Left Panel} --  Modeled linear completeness functions that are perfect (red dotted), moderately underestimated (blue dashed), and moderately overestimated, compared to the ``true'' completeness of the data (grey).  In this example the true completeness is a linear from from 2 to 4 \msun, with a value of 0\% below 2 \msun\ and 100\% above 4 \msun.    {\it Center Panel} -- A histogram of the data (grey) with the ``true'' completeness function applied.  The case of perfectly modeled completeness is in red.  A moderate underestimate of the completeness (blue) indicates that more lower stars are present than expected, whereas a moderate overestimate of the true completeness (green) indicates a deficit of lower mass stars relative to observations.   {\it Right Panel} -- A series of histograms indicating the recovered median values of $\alpha$ for each of the modeled completeness functions.  The distribution of median values is derived from 100 different datasets, each with 10$^{5}$ stars and $\Mmaxinput =$ 60 \msun.  The solid colored lines indicate the median of each distribution, and grey-dashed line is the $\alpha_{\mathrm{input}}$.  A well-characterized completeness function (red) results in an accurate recovery of $\alpha$, whereas a modest overestimate (green) or underestimate (blue) lead to systematic under and overestimates of $\alpha$.}
\label{fig:compperfect}
\end{center}
\end{figure}

Keeping the same true completeness function, we then explored MF recovery for various assumed completeness functions.   In Figure \ref{fig:compperfect}, we considered the case in which the true (grey) and assumed (red) completeness functions were identical linear ramps between 2 and 4 \msun.  We simulated 100 datasets each with 10$^{5}$ stars and $\Mmaxinput =$ 60 \msun, and measured the median value of $\alpha_{\mathrm{recovered}}$, whose distribution is plotted as the red histogram in the right panel. The median of this distribution is represented by the solid red line.  For a perfectly known completeness function, we found excellent recovery of the MF slope, as expected.

We next considered the cases where the true and assumed completeness functions were not identical.  In the first case, the ramp in the assumed completeness function occurred between $M_{\mathrm{Cmin}} =$ 2 and $M_{\mathrm{Cmax}} =$ 3 \msun\ (green), meaning the data are assumed to be more complete actually are, i.e., the stars between 2 and 4 \msun\ are assigned artificially high weights in Equation \ref{eq:comppdf}.  As before, we simulated 100 different clusters of 10$^{5}$ stars, and examined the distribution of the median recovered values of $\alpha$.  As shown in the right-hand panel of Figure \ref{fig:compperfect}, the recovered values of MF slope are clustered around $\alpha =$ 2.0, with small scatter.  This test indicates that a moderate overestimation of the true completeness function can induce a significant underestimate of $\alpha$, because stars that are ``missing'' from the lower mass end due to observational completeness are instead interpreted as being underproduced by a flatter MF.    This finding is similar to that in \citet{asc09}, who demonstrates that an artificial flattening of the MF due to completeness considerations mimics the behavior of predicted mass segregation effects.  We discuss this point further in \S \ref{sec:caveats}.

We also considered a case of a modest underestimation of the true completeness function.  Here, the mass limits on the assumed completeness function were $M_{\mathrm{Cmin}} =$ 2 \msun\ and $M_{\mathrm{Cmax}} =$ 6 \msun.  In this scenario, stars between 2 and 6 \msun\ are assigned artificially low weights in Equation \ref{eq:comppdf}.  Application of this assumed completeness function results in median recovered values of $\alpha =$ 2.7 (blue in Figure \ref{fig:compperfect}).  This exercise indicates that even a moderate underestimation of the true completeness function can induce a significant overestimate of $\alpha$.  

\begin{figure}[t!]
\begin{center}
\plotone{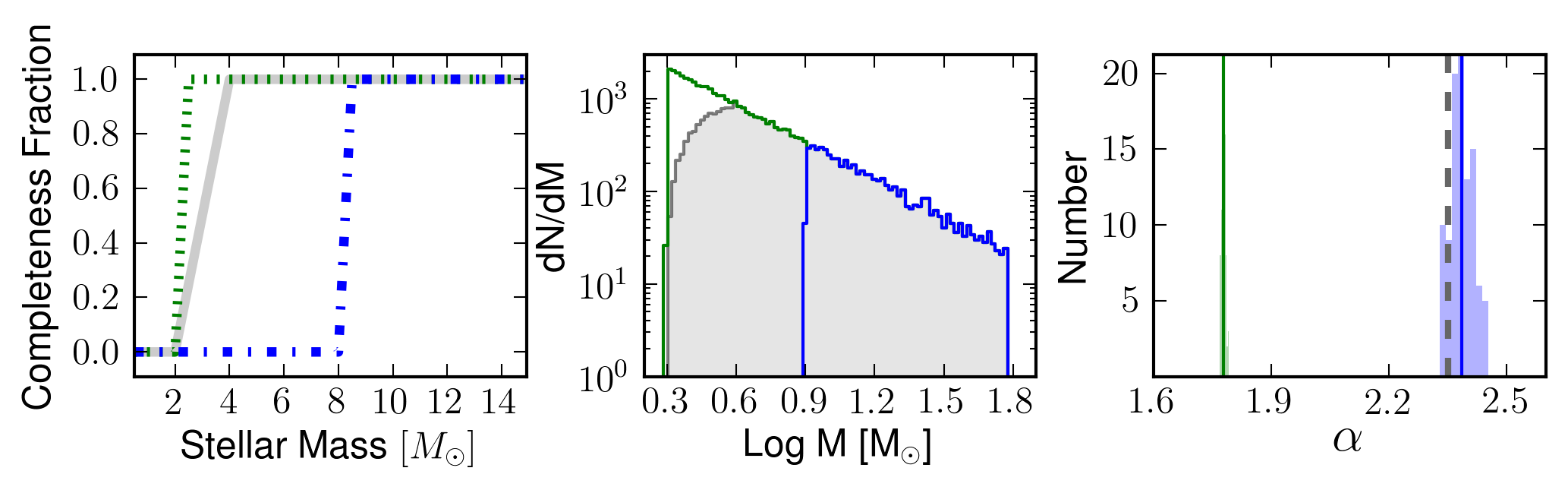}
\caption{The same as Figure \ref{fig:compperfect} only for extreme completeness examples, i.e., all data are incorrectly assumed to be complete (green) and a conservative completeness cut (blue).  The ``true'' completeness function (grey) is a linear ramp from 2 to 4 \msun.  In this case, the assumption that all the data are complete leads to a large deficit of lower mass stars (green, center panel), resulting in an severe systematic underestimate of $\alpha$ (green right panel).  A lack of appropriate treatment of completeness has significant implications for understanding mass segregation in cluster \citep[e.g.,][]{asc09}.  In contrast, a conservative completeness cut (i.e., a step function at 8 \msun), permits an unbiased recovery of the MF slope, at the expense of increased scatter due to the smaller dynamic mass range and decreased number of stars.}
\label{fig:comp23}
\end{center}
\end{figure}

Some MF studies in the literature either apply a conservative completeness limit, i.e., only consider data that are likely to be 100\% complete, or make no attempt to correct for completeness.  We explore the effects of each of these scenarios and present the results in Figure \ref{fig:comp23}.  In the case that no attempt is made to account for completeness (green), i.e., all data above 3 \msun\ are complete, the recovered slope of the MF is $\sim$ 1.7, which is significantly flatter than the true MF slope of 2.35. 

The scenario in which a conservative completeness cut is applied is more promising (blue in Figure \ref{fig:comp23}).  Here, we see that the input value of $\alpha$ is accurately recovered, albeit with lower precision as both the number of stars and dynamic mass range are smaller.

In this section, we have shown that by correctly characterizing completeness, through an example with a linear completeness function, the MF slope can be recovered without bias.  However, failure to include well-characterized completeness corrections can lead to systematic biases.  This finding can readily be generalized to more complex completeness functions, which simply involves replacing $\pobsd$ by another functional or tabulated form.  To determine a completeness function in practice, we strongly advise that extensive artificial star tests should be performed.  Although this process can be a computationally expensive, it is essential to have a well-characterized completeness function in order to accurately measure the MF slope.

\section{Toward Full Forward Modeling of a Cluster's MF}
\label{sec:caveats}

Throughout this paper, we have employed a simple model of the IMF, and not considered other physical (e.g., dynamics) and observational effects (e.g., unresolved binaries), which can influence the accurate measurement of a MF from resolved stars.  Here, we briefly discuss how such outstanding issues can potentially be incorporated into the context of a probabilistic framework for a more holistic model of a cluster's MF.

\textit{(i) Unresolved Binaries --} It is well-established in the literature that many stars have equal or lower mass companions, with binary fractions ranging from $\sim$ 0.35 to 1.0 for massive O stars to $\sim$ 0.3 to 0.5 for solar mass stars \citep[e.g.,][]{pre99, zin07, san11, san12, kim12}.  Such binaries are typically difficult to detect, particularly with photometry, and are frequently treated as single stars.  Several studies have shown that failing to account for unresolved binaries can cause systematic steepening to a cluster's MF slope by $\sim$ 0.1 - 0.5 depending on the intrinsic MF slope \citep[e.g.,][]{sag91, mai08}

In this paper, we have assumed that all data points are from single stars.  However, there are at least two strategies for incorporating binaries into the presented framework.  The first is to construct a model of the probability that a star is a binary based on measured luminosity and color, or estimated mass, of the primary star.  It would be straightforward to employ such a model as a prior on the the mass PDF of each star.  

A second method would be to build binary star tracks into the process of photometric SED fitting, allowing one to compute the probability of a star being better described as a binary, in which case the mass PDF of the primary star would likely be changed dramatically, i.e., it would be flatter as one would expect little constraint on the mass of the primary solely from broadband photometric observations.  The net effect of either approach is to increase the uncertainty in the mass of the primary star, which reduces its contribution to constraining a cluster's MF.  

\textit{ (ii) Mass Segregation --}  Various observations have shown that massive stars are typically located in the central regions of young clusters, while lower mass stars are preferentially found in the outskirts of a cluster.  The origins of observed mass segregation are often attributed either to primordial star formation or internal cluster dynamics (see \citealt{zin07} and \citealt{por10} and references therein).  The general effects of mass segregation on MF measurements include flatter slopes in the inner regions, i.e., the stellar populations have a higher percentage of high mass stars, and a steeper slope in the outskirts, where there are fewer high mass stars.  Mass segregation can also introduce bias into the global MF measurement, depending on the treatment of completeness and the radius to which the data are considered.

Interestingly, simulations conducted by \citet{asc09} have shown that the effects of mass segregation on the MF slope are almost entirely degenerate with observational completeness.  That is, the completeness function of a resolved cluster is a strong function of location such that the completeness limits are brighter in the center, and fainter in the outer regions.  As a result, the recovery of a flatter MF slope in the central region of a cluster may simply be due to not properly correcting for a spatially variable completeness function such that the number of unobserved lower mass stars is significantly underestimated.  

With the presented probabilistic framework, the most straightforward way to account for mass segregation effects in modeling a cluster's global MF is to ensure that the completeness function is well-characterized at all positions within the cluster.  Given that completeness mimics the behavior of mass segregation, an accurate accounting of spatial completeness will mitigate biases on the recovery of a cluster's global MF slope.  At a fundamental level, a well-characterized completeness function necessitates extensive artificial star tests, which can be computationally expensive.   Within the PHAT program, large parts of the fundamental science require extensive such artificial star tests (see \citealt{dal12}).  As such, we anticipate having well-sampled spatially varying completeness functions, which will provide the necessary characterization of spatial variations in the cluster populations to avoid strong biases in the recovered global MF slope.

\textit{(iii) Non-Coeval Populations  --}  It is possible that young clusters are not purely single age, and instead have formed over some finite time interval.  Applying a model that assumes coevality to a non-coeval population can lead to strong systematic biases in the recovered MF slope \citep[e.g.,][]{mil79, sca98, elm06}, particularly when the duration of SF is comparable to the age of the cluster.

The incorporation of an extended SFH into our probabilistic framework is fairly straightforward.  It requires multiplying the intrinsic MF by an integral of a parameterized star formation history (cf. Equation 1 in \citealt{elm06}), and then using MCMC sampling to simultaneously constrain the cluster's MF slope and star formation history, given the data.  This approach forms the basis of color-magnitude diagram modeling techniques such as those discussed in \citet{dol02} and \citet{har01}, and is known to lead to very broad MF slope constraints for objects with extended SFHs, e.g., galaxies \citep[e.g.,][]{wei11}.  

\textit{(iv) Cluster Membership --} In practice, there is always some degree of ambiguity as to whether a given star is a member of the cluster or whether it is a member of the surrounding field population.  Thus, it is necessary to assign a membership probability to each star.  Within the presented framework, the membership probability enters as a linear multiplication term that simply serves to weight each $p(\vec{d} \midline \vtheta, \mathrm{obs})$, where the weighting factor ranges from zero (not a cluster member) to unity (definitely a cluster member).  The simplest way to assess the probability of cluster membership is through a statistical comparison with a color-magnitude diagram of a surrounding field population.  

\section{Conclusions}
\label{sec:conclude}

We have presented a probabilistic approach to constrain the MF of a young resolved stellar cluster.  This framework allows the incorporation of uncertainties in the masses of individual stars that may arise from the conversion of observed fluxes into stellar masses,  avoids binning the measurement in mass, explicitly deals with completeness functions that may not be a simple step function of stellar mass, and assigns meaningful error bars to the parameters of interest.

Adopting a single-sloped power-law MF model, we explored the ability of this approach to constrain MF parameters using mock data.  In particular, we found:

\begin{itemize}
\item For highly idealized mock datasets (perfectly known masses and completeness), we recover the slope of the input IMF with no systematic biases, and to a precision that depends on the number of observed stars and the dynamic range of the observed masses, i.e., $\log(\Mmaxobs/\Mcomp)$.  We verified that one can only derive a lower limit estimate on $\Mmax$, namely $\Mmaxobs$, which is not particularly informative in terms of understanding the upper mass limit of the IMF.\\

\item We computed the theoretical precision, i.e., lower limit, to which $\alpha$ can be measured as a function of the observed mass range, $\log(\Mmaxobs/\Mcomp)$ and the number of stars, $N$, which we approximate analytically as a 3rd order polynomial. We compared the theoretical precision with selected literature IMF studies and found that $\sim$ 3/4 of the literature studies quoted error bars below the theoretical limit, usually by a factor of $\sim$ 2.  After assigning these studies new and larger error bars based on the theoretical limit, we computed the weighted mean and weighted 1-$\sigma$ values from the literature studies and found $\langle \alpha \rangle =$ 2.46 with a 1-$\sigma$ dispersion of 0.35 dex.  The broad uncertainties indicate that the mean $\alpha$ is consistent with several common IMFs such as Salpeter, Kroupa, Kennicutt, and Scalo, suggesting that the current state of MF studies have little leverage on discerning between significantly different high mass IMF slopes.  This finding reinforces the need for a large scale systematic study of the high mass IMF in order to provide the much needed empirical constraints.  \\

\item We then considered the case where the masses are not perfectly known.   Specifically, we generated mock mass lists where each mass had a log-normal error distribution.  Using the same datasets, we demonstrated the differences in MF recovery using models that did and did not account for the effects of uncertain masses.  In general, we found that if all the mass uncertainties were small, the recovery of $\alpha$ was not substantially biased.  However, in the case of intermediate to large mass uncertainties, a failure to model the mass uncertainties resulted in systematic biases in the slope of recovered MF, the magnitude and direction of which depend on the magnitude of the uncertainty, e.g., moderate uncertainties result in too steep of a slope, while large errors result in too flat of a slope.  Similarly, not accounting for mass uncertainties led to severe systematic overestimates of $\Mmaxinput$ (or underestimates of a cluster's true age).  In contrast, applying models that account for uncertainties in the stellar masses resulted in near-perfect recovery of the input IMF slope with precision that was comparable to the case of perfectly known masses, in all except the cases with the most extreme mass uncertainties.  We found constraints on $\Mmax$ to be in agreement with those from the case of highly idealized data.\\

\item For a completeness function that grows linearly from 0 to 1 within a certain mass range, we found that perfect knowledge of the completeness function resulted in excellent recovery of the MF slope.  However, both moderately and extremely incorrect completeness functions led to strong systematic biases in the recovered MF slope.  We strongly recommend that extensive artificial stars tests be used in all resolved MF studies.  In lieu of this option, only data that are nearly 100\% complete should be utilized to minimize systematic biases, although the precision on the MF constraints will decrease in this case.\\

\item We discussed factors that can influence MF slope determination, but that were not included in this paper.  Such effects include unresolved binaries, mass segregation, non-coevality, and cluster membership.  In each case, we suggest ways in which such uncertainties can be folded into the presented probabilistic framework.

\end{itemize}

\acknowledgments

The authors are grateful for the thorough, critical, and extremely insightful referee's report from John Scalo.  His comments greatly improved the clarity and accessibility of the paper.   We would also like to thank Nate Bastian, Mario Gennaro, and Andrea Stolte for interesting discussions on the IMF and Phil Marshall for insightful comments on improving visualizations.  D.G. kindly acknowledges financial support by the German Research Foundation (DFG) through grant GO~1659/3-1. Support for this work was provided by NASA through grant number HST GO-12055 from the Space Telescope Science Institute, which is operated by AURA, Inc., under NASA contract NAS5-26555. This research made extensive use of NASA's Astrophysics Data System Bibliographic Services.

\appendix
\label{sec:appendix}

\section{Markov chain Monte Carlo}

Markov chain Monte Carlo (MCMC) algorithms are a broad class of numerical techniques for estimating the form of probability
distributions. There are many textbooks \citep{bishop-book, nr-book, mackay-book, gelman-book} that describe the general formalism and application of various MCMC methods. Here, we provide a brief  overview of the general theory behind MCMC comment on the specific algorithm that we employ in this paper. We refer the interested reader to the above references for more detailed information.

MCMC provides a prescription for drawing a set of unbiased samples from a probability distribution function (PDF) that can be \emph{evaluated} (up to a normalization constant) given a set of parameters. In this paper, we use MCMC to estimate the distribution of model parameters ($\alpha$, $M_\mathrm{max}$, and $N_\mathrm{pred}$) for a cluster's mass function (MF) that is consistent with the observations and marginalized over the nuisance parameters---$M_\mathrm{max}$ and $N_\mathrm{pred}$---when only the MF slope, $\alpha$, is desired, for example. After running a MCMC chain and obtaining $K$ samples $\theta_k = \{\alpha_k, M_{\mathrm{max}, k}, N_{\mathrm{pred},k}\}$, the expectation value of $\alpha$ conditioned on the data and marginalized over $\theta_{-\alpha} = \{M_\mathrm{max}, N_\mathrm{pred}\}$ is approximately given by

\begin{equation}
    \left < \alpha \right > = \int \alpha \, p(\theta \midline \{M_i\}, N) \, \mathrm{d}\theta_{-\alpha} \sim  \frac{1}{N} \, 
    \sum_{n=1}^N \alpha_k \quad.
\end{equation}

Similarly, the marginalized PDF for $\alpha$
\begin{equation}
    p(\alpha \midline \{ M_i \}, N) = \int p(\theta \midline
    \{M_i\}, N) \, \mathrm{d} \theta_{-\alpha} \, ,
\end{equation}
is given by the \emph{histogram} of samples projected into the $\alpha$ plane.  This marginalized PDF provides the desired constraint on the MF slope, while accounting for degeneracies with nuisance parameters. 

MCMC is generally much more efficient than alternatives such as rejection
sampling or grid-based methods because it requires fewer calculations to
provide a representative sampling of the density.

The most commonly used MCMC algorithm is called the Metropolis-Hastings (M--H)
algorithm \citep{met53, has70}. While this is not the algorithm we use, it is instructive to
outline how it works. The M--H algorithm ``walks'' around the parameter space
stochastically starting from a user-defined initial position. Each step in the
chain is determined by sampling a proposal position $\theta^\prime$ from a
(simple) distribution $Q(\theta^\prime \midline \theta^{(t)})$ that only depends on
the \emph{current} position $\theta^{(t)}$ and then accepting (with
replacement) this position with the probability
\begin{equation}
    A(\theta^\prime ; \theta^{(t)}) = \mathrm{min} \left \{
    1, \, \frac{p(\theta^\prime|\{M_i\},N)}{p(\theta^{(t)} \midline \{M_i\},N)} \,
    \frac{Q(\theta^{(t)} \midline \theta^\prime)}{Q(\theta^\prime \midline \theta^{(t)})}
    \right \} \quad.
\end{equation}

For this project, we used an affine-invariant ensemble algorithm
\citep[\texttt{emcee}\footnote{\url{http://danfm.ca/emcee}};][]{for12, goo10}.
This algorithm is more efficient than M--H when sampling a non-trivial
density and it requires significantly less fine-tuning to achieve good
performance. Instead of exploring the parameter space sampling a single point
at a time, \texttt{emcee} includes an ensemble of $L$ coupled ``walkers'' and
the proposal distribution $Q(\theta^\prime_\ell \midline \theta_{-\ell})$ for a
particular walker $\ell$ is based on the current positions of all the walkers.
Specifically, an update step involves randomly choosing a walker from the
complimentary set $\theta_j \in \theta_{-\ell} = \{ \theta_1^{(t+1)}, \cdots,
\theta_{\ell-1}^{(t+1)}, \theta_{\ell+1}^{(t)}, \cdots, \theta_L^{(t)} \}$
and proposing a position along the vector between the two walkers
\begin{equation}
    \theta^\prime = \theta_j + Z \, \left [
        \theta_\ell^{(t)} - \theta_j
    \right ]
\end{equation}
where $Z$ is a random variable distributed according to
\begin{equation}
    p(Z) \propto \left \{ \begin{array}{ll}
        {\displaystyle \frac{1}{\sqrt{Z}}} & \mathrm{if}\,
            {\displaystyle \frac{1}{a}} < Z < a \\
        0 & \mathrm{otherwise}
    \end{array}\right .
\end{equation}
for some $a > 1$. The proposal $\theta_\ell^\prime$ is then accepted (again
with replacement) with the probability
\begin{equation}
    A(\theta^\prime_\ell ; \theta^{(t)}_\ell) = \mathrm{min} \left \{
    1, \, Z^{D-1} \, \frac{p(\theta^\prime_\ell|\{M_i\},N)}
        {p(\theta^{(t)}_\ell \midline \{M_i\},N)}
    \right \}
\end{equation}
where $D$ is the dimension of the parameter space.

\clearpage

\LongTables
\begin{deluxetable*}{lccccccccc}
\tablecolumns{10}
\tablecaption{Literature Constraints on the IMF Slope}
\tablehead{
   \colhead{} &
    \colhead{Cluster} &
    \colhead{$\log(N)$} &
    \colhead{$\log(\Mmaxobs/\Mcomp)$} &
    \colhead{$\alpha$} &
    \colhead{$\sigma$} &
    \colhead{Min. Mass} &
    \colhead{Max. Mass} &
     \colhead{$\frac{\Delta \alpha_{literature}}{\Delta \alpha_{theory}}$} &
     \colhead{Reference} \\
      \colhead{} &
  \colhead{Name} &
    \colhead{} &
    \colhead{} &
    \colhead{} &
      \colhead{} &
    \colhead{(M$_{\odot}$)} &
    \colhead{(M$_{\odot}$)} &
              \colhead{}  &
               \colhead{} \\
  \colhead{(1)} &
    \colhead{(2)} &
    \colhead{(3)} &
    \colhead{(4)} &
    \colhead{(5)} &
    \colhead{(6)} &
    \colhead{(7)} &
     \colhead{(8)} &
      \colhead{(9)} &
       \colhead{(10)} \\
}
\startdata 
1	& NGC2323 &	3.31 & 0.99 & 2.94 & 0.15 & 0.4 & 3.9 & 1.66	& Kalirai et al.~2003\\ 
2	& M11 &	3.26 & 0.49 & 2.49 & 0.09 & 1.1 & 3.4 & 0.32	& Santos et al.~2005\\ 
3	& NGC663 &	 3.15 & 0.8 & 2.38 & 0.22 & 1.6 & 10.0 & 1.54	& Pandey et al.~2005\\ 
4	& NGC2168 &	3.0 & 0.78 & 2.29 & 0.27 & 0.6 & 3.6 & 1.76	& Kalirai et al.~2003\\ 
5	& NGC2422 &	2.6 & 0.44 & 3.07 & 0.08 & 0.9 & 2.5 & 0.23	& Prisinzano et al.~2003\\ 
6	& Tr23 &	2.48 & 0.32 & 2.4 & 0.6 & 1.0 & 2.1 & 1.42	& Bonatto et al.~2007\\ 
7	& Stock2 &	2.31 & 0.44 & 3.01 & 0.4 & 1.5 & 4.1 & 1.05	& Sanner et al.~2001\\ 
8	& NGC4852 &	2.3 & 0.73 & 2.3 & 0.3 & 0.6 & 3.2 & 1.38	& Carraro et al.~2005\\ 
9	& NGC654 &	2.26 & 0.98 & 2.16 & 0.05 & 1.2 & 11.5 & 0.41	& Pandey et al.~2005\\ 
10	& NGC4349 &	2.21 & 0.44 & 2.18 & 0.12 & 1.8 & 5.0 & 0.3	& Tarrab et al.~1982\\  
11	& Orion &	 2.13 & 1.04 & 2.39 & 0.15 & 1.3 & 14.1 & 1.31	& Tarrab et al.~1982\\  
12	& Pleiades &	2.1 & 0.61 & 2.99 & 0.4 & 1.0 & 4.1 & 1.31	& Sanner et al.~2001\\ 
13	& Ly9 &	2.09 & 0.32 & 2.1 & 0.7 & 1.1 & 2.3 & 1.41	& Bonatto et al.~2007\\ 
14	& NGC3532 &	2.06 & 0.45 & 2.55 & 0.21 & 1.6 & 4.5 & 0.5	& Tarrab et al.~1982\\  
15	& Ly4 &	2.02 & 0.28 & 2.3 & 0.2 & 1.0 & 1.9 & 0.37	& Bonatto et al.~2007\\ 
16	& NGC5715 &	1.97 & 0.3 & 2.3 & 0.5 & 1.1 & 2.2 & 0.92	& Bonatto et al.~2007\\ 
17	& CygOB2 &	1.97 & 1.0 & 1.9 & 0.2 & 10.0 & 100.0 & 1.48	& Massey et al.~1995\\ 
18	& $\alpha$Per &	1.95 & 0.84 & 2.73 & 0.23 & 1.3 & 8.9 & 1.14	& Tarrab et al.~1982\\  
19	& Pleiades &	1.94 & 0.69 & 2.89 & 0.27 & 1.3 & 6.3 & 0.93	& Tarrab et al.~1982\\  
20	& LH9 &	1.92 & 0.74 & 2.4 & 0.2 & 10.0 & 55.0 & 0.78	& Massey et al.~1995\\ 
21	& NGC346 &	1.92 & 0.85 & 2.3 & 0.1 & 10.0 & 70.0 & 0.5	& Massey et al.~1995\\ 
22	& Tr14/16 &	1.91 & 1.08 & 2.0 & 0.2 & 10.0 & 120.0 & 1.72	& Massey et al.~1995\\ 
23	& LH58 &	1.82 & 0.7 & 2.4 & 0.2 & 10.0 & 50.0 & 0.66	& Massey et al.~1995\\ 
24	& Praesepe &	1.81 & 0.39 & 3.12 & 0.39 & 1.3 & 3.2 & 0.72	& Tarrab et al.~1982 \\ 
25	& LH10 &	1.81 & 0.95 & 2.1 & 0.1 & 10.0 & 90.0 & 0.6	& Massey et al.~1995\\ 
26	& NGC2516 &	1.81 & 0.65 & 2.95 & 0.33 & 2.0 & 8.9 & 0.97	& Tarrab et al.~1982\\  
27	& NGC6067 &	1.8 & 0.5 & 3.18 & 0.39 & 4.0 & 12.6 & 0.85	& Tarrab et al.~1982\\  
28	& Haydes &	1.79 & 0.33 & 3.58 & 0.48 & 1.3 & 2.8 & 0.82	& Tarrab et al.~1982\\  
29	& NGC2264 &	1.78 & 0.56 & 2.53 & 0.29 & 2.2 & 7.9 & 0.69	& Tarrab et al.~1982\\  
30	& NGC457 &	1.71 & 0.45 & 1.36 & 0.05 & 5.0 & 14.1 & 0.09	& Tarrab et al.~1982\\  
31	& NGC6405 &	1.68 & 0.65 & 2.84 & 0.37 & 2.0 & 8.9 & 0.99	& Tarrab et al.~1982\\  
32	& NGC2281 &	1.67 & 0.6 & 4.17 & 0.53 & 1.4 & 5.6 & 1.27	& Tarrab et al.~1982\\  
33	& NGC6281 &	1.63 & 0.45 & 2.14 & 0.23 & 1.6 & 4.5 & 0.4	& Tarrab et al.~1982\\  
34	& NGC6633 &	1.62 & 0.29 & 3.08 & 0.45 & 1.8 & 3.5 & 0.66	& Tarrab et al.~1982 \\ 
35	& Cz37	& 1.61 & 0.14 & -0.1 & 0.9 & 1.8 & 2.5 & 1.27	& Bonatto et al.~2007\\ 
36	& LH117/118 &	1.6 & 1.0 & 2.6 & 0.2 & 10.0 & 100.0 & 1.08	& Massey et al.~1995\\ 
37	& NGC2099 &	1.59 & 0.51 & 2.6 & 0.38 & 2.2 & 7.1 & 0.71	& Tarrab et al.~1982\\  
38	& NGC4755 &	1.58 & 0.4 & 2.1 & 0.23 & 5.6 & 14.1 & 0.36	& Tarrab et al.~1982\\  
39	& NGC6025 &	1.57 & 0.45 & 3.06 & 0.5 & 2.5 & 7.1 & 0.84	& Tarrab et al.~1982\\  
40	& NGC2362 &	1.56 & 0.75 & 2.3 & 0.32 & 2.5 & 14.1 & 0.96	& Tarrab et al.~1982\\  
41	& NGC2451 &	1.54 & 0.78 & 2.1 & 0.27 & 1.3 & 7.9 & 0.86	& Tarrab et al.~1982\\  
42	& IC2602 &	1.53 & 0.63 & 1.88 & 0.21 & 1.3 & 5.6 & 0.48	& Tarrab et al.~1982\\  
43	& NGC7243 &	1.48 & 0.41 & 1.83 & 0.18 & 2.2 & 5.6 & 0.26	& Tarrab et al.~1982\\  
44	& IC4665 &	1.48 & 0.54 & 1.73 & 0.16 & 1.8 & 6.3 & 0.29	& Tarrab et al.~1982\\  
45	& NGC6611 &	1.48 & 0.88 & 1.9 & 0.2 & 10.0 & 75.0 & 0.71	& Massey et al.~1995\\ 
46	& NGC1960 &	1.4 & 0.41 & 1.24 & 0.05 & 3.5 & 8.9 & 0.07	& Tarrab et al.~1982\\  
47	& NGC5662 &	1.4 & 0.6 & 1.76 & 0.19 & 2.5 & 10.0 & 0.35	& Tarrab et al.~1982\\  
48	& NGC2422 &	1.38 & 0.65 & 2.38 & 0.42 & 2.0 & 8.9 & 0.84	& Tarrab et al.~1982\\  
49	& NGC1662 &	1.38 & 0.44 & 2.16 & 0.32 & 1.8 & 5.0 & 0.44	& Tarrab et al.~1982\\  
50	& NGC5460 &	1.38 & 0.35 & 3.35 & 0.71 & 2.5 & 5.6 & 0.88	& Tarrab et al.~1982\\  
51	& IC1805 &	1.38 & 1.0 & 2.3 & 0.2 & 10.0 & 100.0 & 0.72	& Massey et al.~1995\\ 
52	& NGC2539 &	1.36 & 0.2 & 2.61 & 0.39 & 2.0 & 3.2 & 0.44	& Tarrab et al.~1982\\  
53	& NGC7092 &	1.36 & 0.29 & 1.72 & 0.16 & 1.8 & 3.5 & 0.19	& Tarrab et al.~1982\\  
54	& NGC3766 &	1.36 & 0.5 & 1.53 & 0.12 & 4.5 & 14.1 & 0.18	& Tarrab et al.~1982\\  
55	& NGC2548 &	1.36 & 0.3 & 4.05 & 0.94 & 2.0 & 4.0 & 1.11	& Tarrab et al.~1982\\  
56	& NGC6823 &	1.36 & 0.6 & 2.3 & 0.4 & 10.0 & 40.0 & 0.71	& Massey et al.~1995\\ 
57	& IC2391 &	1.36 & 0.65 & 2.13 & 0.34 & 1.4 & 6.3 & 0.67	& Tarrab et al.~1982\\  
58	& NGC2301 &	1.34 & 0.6 & 2.56 & 0.49 & 1.4 & 5.6 & 0.85	& Tarrab et al.~1982\\  
59	& NGC884 &	1.34 & 0.5 & 1.29 & 0.06 & 4.5 & 14.1 & 0.09	& Tarrab et al.~1982\\  
60	& NGC7160 &	1.34 & 0.8 & 2.52 & 0.45 & 2.0 & 12.6 & 1.14	& Tarrab et al.~1982\\  
61	& NGC4609 &	1.32 & 0.55 & 2.02 & 0.3 & 2.8 & 10.0 & 0.46	& Tarrab et al.~1982\\  
62	& Cr140 &	1.32 & 0.6 & 2.58 & 0.51 & 2.0 & 7.9 & 0.85	& Tarrab et al.~1982\\  
63	& NGC2571 &	1.3 & 0.66 & 3.3 & 0.65 & 2.2 & 10.0 & 1.19	& Tarrab et al.~1982\\  
64	& NGC2439 &	1.3 & 0.3 & 1.6 & 0.14 & 7.1 & 14.1 & 0.15	& Tarrab et al.~1982\\  
65	& NGC1893 &	1.28 & 0.81 & 2.6 & 0.3 & 10.0 & 65.0 & 0.7	& Massey et al.~1995\\ 
66	& IC2581 &	1.26 & 0.64 & 2.07 & 0.36 & 3.2 & 14.1 & 0.61	& Tarrab et al.~1982\\  
67	& NGC1528 &	1.26 & 0.45 & 1.15 & 0.03 & 2.0 & 5.6 & 0.04	& Tarrab et al.~1982\\  
68	& NGC2482 &	1.26 & 0.51 & 4.33 & 0.95 & 2.2 & 7.1 & 1.26	& Tarrab et al.~1982\\  
69	& NGC2251 &	1.2 & 0.49 & 2.36 & 0.49 & 1.6 & 5.0 & 0.6	& Tarrab et al.~1982\\  
70	& NGC6242 &	1.2 & 0.5 & 2.15 & 0.39 & 4.0 & 12.6 & 0.48	& Tarrab et al.~1982\\  
71	& NGC2232 &	1.2 & 0.6 & 2.73 & 0.63 & 2.0 & 7.9 & 0.92	& Tarrab et al.~1982\\  
72	& NGC6531 &	1.2 & 0.76 & 1.94 & 0.34 & 2.5 & 14.5 & 0.66	& Tarrab et al.~1982\\  
73	& NGC1664 &	1.18 & 0.26 & 3.62 & 0.96 & 2.5 & 4.5 & 0.9	& Tarrab et al.~1982\\  
74	& NGC581 &	1.18 & 0.35 & 3.09 & 0.78 & 6.3 & 14.1 & 0.78	& Tarrab et al.~1982\\  
75	& NGC7790 &	1.15 & 0.45 & 2.44 & 0.54 & 4.5 & 12.6 & 0.58	& Tarrab et al.~1982\\  
76	& NGC3590 &	1.15 & 0.4 & 2.07 & 0.36 & 5.6 & 14.1 & 0.37	& Tarrab et al.~1982\\  
77	& NGC6709 &	1.15 & 0.2 & 1.98 & 0.28 & 3.5 & 5.6 & 0.25	& Tarrab et al.~1982\\  
78	& NGC6871 &	1.11 & 0.35 & 2.18 & 0.4 & 5.6 & 12.6 & 0.38	& Tarrab et al.~1982\\  
79	& NGC5281 &	1.11 & 0.46 & 2.02 & 0.36 & 3.5 & 10.0 & 0.38	& Tarrab et al.~1982\\  
80	& NGC2169 &	1.08 & 0.6 & 1.74 & 0.27 & 2.8 & 11.2 & 0.35	& Tarrab et al.~1982\\  
81	& NGC1502 &	1.08 & 0.4 & 1.56 & 0.18 & 5.6 & 14.1 & 0.17	& Tarrab et al.~1982\\  
82	& NGC1342 &	1.08 & 0.36 & 3.46 & 1.05 & 2.2 & 5.0 & 0.96	& Tarrab et al.~1982\\  
83	& NGC2244 &	1.08 & 0.85 & 1.8 & 0.3 & 10.0 & 70.0 & 0.55	& Massey et al.~1995\\ 
84	& NGC6871 &	1.04 & 0.6 & 1.9 & 0.4 & 10.0 & 40.0 & 0.51	& Massey et al.~1995\\ 
85	& NGC7380 &	1.04 & 0.81 & 2.7 & 0.3 & 10.0 & 65.0 & 0.51	& Massey et al.~1995\\ 
86	& Berkeley86 & 	1.0 & 0.6 & 2.7 & 0.4 & 10.0 & 40.0 & 0.49	& Massey et al.~1995\\ 
87	& NGC2323 &	0.9 & 0.2 & 4.42 & 1.73 & 5.0 & 7.9 & 1.36	& Tarrab et al.~1982 \\ 
88	& NGC6913 &	0.78 & 0.6 & 2.1 & 0.6 & 10.0 & 40.0 & 0.72	& Massey et al.~1995\\ 
89	& CepOB5 &	0.78 & 0.48 & 3.1 & 0.6 & 10.0 & 30.0 & 0.59	& Massey et al.~1995
\enddata
\tablecomments{\small The literature IMF studies used in this paper.  We have tabulated the number of stars (3) over the given mass range (4), and the reported values of $\alpha$ (5), where all values have been updated to reflect our usage of $\alpha_{Salpeter} =$ 2.35 and the 1-$\sigma$ uncertainty on $\alpha$ (6), allowing for a direct comparison with our calculated theoretical lower limits. The values in column (9) have been computed using the 1-$\sigma$ value listed in the literature, i.e., column (6) in this table, and the theoretical precision presented in Figure \ref{fig:litcompabs}.  There, we also show that nearly $\sim$ 3/4 of the literature considered quote error bars smaller than the theoretical lower limit.}
\label{tab1}
\end{deluxetable*}

\clearpage

\begin{deluxetable*}{cc}
\tablecolumns{2}
\tablecaption{Polynomial Coefficients for $\Delta \alpha$ Approximation}
\tablehead{
  \colhead{i,j} &
    \colhead{$\beta_{i,j}$} \\
}
\startdata 
0,0 & 2.90 \\
0,1 & -2.58 \\
0,2 & 1.00 \\
0,3 & -0.25 \\
1,0 & -0.49 \\
1,1 & -3.51 \\
1,2 & 5.00 \\
1,3 & -1.78 \\
2,0 & -0.05 \\
2,1 & 1.30 \\
2,2 & -1.56 \\
2,3 & 0.52 \\
3,0 & 0.01 \\
3,1 & -0.09 \\
3,2 & 0.09 \\
3,3 & -0.02
\enddata
\tablecomments{Coefficients for the 3rd order polynomial approximation of the theoretical precision on the IMF slope, $\Delta \alpha$, as a function of the observed number of stars and observed mass range (see Equation \ref{eq:deltaalpha}).}
\label{tab2}
\end{deluxetable*}

\end{document}